\documentclass[14pt]{article}
\pdfoutput=1
\usepackage{amsmath,amssymb,amsfonts,amsthm,bm,bbm,cancel,wasysym}
\usepackage{epsfig,graphics,graphicx,epstopdf}
\usepackage{array,booktabs,colortbl,colordvi,multirow}
\usepackage{colordvi,color,xcolor}
\usepackage{hyperref}

\usepackage{rotating}

\usepackage{verbatim}
\usepackage{cite}
\usepackage{subfig}
\usepackage{setspace}
\usepackage{url}
\usepackage[percent]{overpic}
\usepackage{slashed}
\usepackage{authblk}

\usepackage{xspace}
\usepackage{fullpage}
\usepackage{hyperref}

\def\ie{{\it i.e.}}
\def\eg{{\it e.g.}}
\def\etc{{\it etc}}

\def\to{\rightarrow}

\newskip\zatskip \zatskip=0pt plus0pt minus0pt
\def\matth{\mathsurround=0pt}
\def\lsim{\mathrel{\mathpalette\atversim<}}
\def\gsim{\mathrel{\mathpalette\atversim>}}
\def\atversim#1#2{\lower0.7ex\vbox{\baselineskip\zatskip\lineskip\zatskip
  \lineskiplimit 0pt\ialign{$\matth#1\hfil##\hfil$\crcr#2\crcr\sim\crcr}}}

\begin{document}


\begin{flushright}
SLAC-PUB-17779\\
\today
\end{flushright}
\vspace*{5mm}

\renewcommand{\thefootnote}{\fnsymbol{footnote}}
\setcounter{footnote}{1}

\begin{center}

{\Large {\bf Towards UV-Models of Kinetic Mixing and Portal Matter: \\ A More Complex Dark Matter Sector?
}}\\

\vspace*{0.75cm}

{\bf Thomas G. Rizzo}~\footnote{rizzo@slac.stanford.edu}

\vspace{0.5cm}

{SLAC National Accelerator Laboratory}\\ 
{2575 Sand Hill Rd., Menlo Park, CA, 94025 USA}

\end{center}
\vspace{.5cm}


\begin{abstract}
\noindent  

Portal Matter (PM), having both Standard Model (SM) and dark sector charges, can induce kinetic mixing (KM) between the $U(1)_D$ dark photon (DP) and the SM gauge fields at the 1-loop 
level offering an attractive mechanism by which light ($\lsim 1$ GeV) thermal dark matter (DM) can interact with visible matter and obtain its observed relic density. In doing so, if the DM is 
fermionic, the CMB and other astrophysical observations inform us that it must be Majorana/pseudo-Dirac in nature to avoid velocity/temperature-independent $s$-wave annihilation to SM final states.  
How does this idea fit into a more UV-complete picture also including the SM interactions? There are some reasons to believe that at least a first step along this path may not lie too far away in 
energy due to the RGE running of the dark gauge coupling, which for a significant range of parameters, becomes non-perturbative at/before the $\sim 10$'s of TeV energy range. This implies that 
$U(1)_D$ must become embedded in an asymptotically free, non-Abelian group, $G_D$, before this can occur. The breaking of this larger group then produces the masses for the PM and the  
additional gauge fields associated with $G_D$ then can lead to new interactions between the SM and the dark sector. Following several bottom-up approaches, we have examined a set of 
distinctive and testable phenomenological features associated with this general setup, based upon a number of simplifying assumptions. Clearly, it behooves us to explore the impact of these specific 
assumptions on these predictions for the array of possible experimental tests of this class of models. In most past analyses it has been assumed that DM is a vector-like, complex singlet under the 
group $G_D$. If this assumption is relaxed, the dark sector must be augmented by additional fermion(s) and the associated scalar fields needed to break the gauge symmetries while generating 
the needed Majorana-like mass terms for the DM. In this paper, we analyze the simplest extension of this kind wherein the DM lies in a vector-like doublet of $G_D$, which we take to have the 
structure $SU(2)_I\times U(1)_{Y_I}$ as in earlier work, leading to new phenomenological implications. We find, for example, that given the current LHC search constraints on the masses of heavy 
gauge bosons, the production of these new dark states with large rates is unlikely to occur at colliders unless they are produced singly in $gg$-fusion or their pair production cross sections are 
resonantly enhanced. We also find that an additional mechanism arises to generate hierarchal neutrino masses in such a setup.

\end{abstract}

\vspace{0.5cm}
\renewcommand{\thefootnote}{\arabic{footnote}}
\setcounter{footnote}{0}
\thispagestyle{empty}
\vfill
\newpage
\setcounter{page}{1}



\section{Introduction and Overview}

While the Standard Model (SM) has enjoyed enormous phenomenological success, it leaves many questions still unanswered, amongst the most compelling of which is the nature of Dark 
Matter (DM).  As of now, the existence of DM is only known through its gravitational interactions but it is likely that for it to achieve the observed relic density\cite{Planck:2018vyg}, assuming 
that DM is indeed made of fundamental particles, it needs to have additional interactions with (at least some of) the familiar SM species. While historically both QCD 
axions\cite{Kawasaki:2013ae,Graham:2015ouw,Irastorza:2018dyq} and thermal WIMPS in the few GeV to $\sim 100$ TeV mass range\cite{Arcadi:2017kky,Roszkowski:2017nbc,Arcadi:2024ukq} 
have been the most popular possibilities for particle DM, and searches for such states continue on all 
fronts\cite{LHC,Aprile:2018dbl,Fermi-LAT:2016uux,Amole:2019fdf,LZ:2022ufs}, it is now well-known that the set of potential DM candidates covers a 
wide spectrum of both mass and coupling that is rather vast\cite{Alexander:2016aln,Battaglieri:2017aum,Bertone:2018krk,Cooley:2022ufh,Boveia:2022syt,Schuster:2021mlr,Cirelli:2024ssz}. 
The types and varieties of searches needed to cover even a fraction of this huge parameter space are daunting, necessarily employing a growing set of quite novel ideas. However,  
it is also expected that in a substantial fraction of this large space, the new interaction(s) between DM and the SM can be described by a set of effective field theories known as `portals', which 
may or may not be renormalizable depending upon the specific scenario under consideration, but all of which posit the existence of a further new class of particles. 

Of this large set of possibilities, one of the most attractive (and, hence, most popular) is an extension of the thermal WIMP\cite{Steigman:2015hda,Saikawa:2020swg} idea into the few MeV to 
$\sim 1$ GeV mass range which is usually constructed within the renormalizable Kinetic Mixing/vector portal setup\cite{KM,vectorportal,Gherghetta:2019coi}. In such models, the DM carries a 
new `dark charge', $Q_D\neq 0$, realized under a new gauge interaction - most simply a new (dark) $U(1)_D$ - which has an associated gauge boson, the dark photon (DP), $V$
\cite{Fabbrichesi:2020wbt,Graham:2021ggy,Barducci:2021egn}. Under his new $U(1)_D$ group, all of the SM fields are assumed to be neutral, \ie, have $Q_D=0$, and so only the manner by which 
the SM and DM fields interact at the $\sim 1$ TeV scale or below is via the kinetic mixing (KM) of the DP with usually, \eg, the $U(1)_Y$ hypercharge field, $B$, in the SM. 
Since both the DM and the DP are assumed to have masses [the later, \eg, via a generalization of the usual Higgs mechanism employing dark Higgs (DH)\cite{Li:2024wqj} vev(s)] below 
$\sim 1$ GeV in this setup, it is sometimes conventional to consider this KM in the low-energy EFT operating at such scales as that between $V$ and the usual SM photon, $A$. In either 
case, this KM can only be generated if there exists some new heavy fields, which are either complex scalars and/or vector-like fermions
\cite{CarcamoHernandez:2023wzf,CMS:2024bni,Alves:2023ufm,Banerjee:2024zvg,Guedes:2021oqx,Adhikary:2024esf} (to avoid the well-known unitarity, electroweak and Higgs coupling constraints), 
that carry both $U(1)_D$ as well as SM gauge charges (\eg, hypercharge and/or electric charge). Here we refer to such new particles Portal Matter (PM) and they have been the subject of 
much recent study\cite{Rizzo:2018vlb,Rueter:2019wdf,Kim:2019oyh,Rueter:2020qhf,Wojcik:2020wgm,Rizzo:2021lob,Rizzo:2022qan,Wojcik:2022rtk,Rizzo:2022jti,Rizzo:2022lpm,Wojcik:2022woa,Carvunis:2022yur,Verma:2022nyd,Rizzo:2023qbj,Wojcik:2023ggt,Rizzo:2023kvy,Rizzo:2023djp}{\footnote{See also Ref.\cite{Ardu:2024bxg}} 
The existence of these new states is then seen to generate vacuum polarization-like graphs linking the usual photon, $A$ (or the SM hypercharge gauge boson) and a 
DP, $V$, at either end, and which, when the heavy PM states are integrated out, leads to a new, renormalizable dim-4 interaction. In the IR limit, applicable 
at energies far below the weak scale, this KM term in the Lagrangian can be simply written as
\begin{equation}
{\cal L}_{AV}=\frac{\epsilon}{2}F_{\mu\nu}^AF^{\mu\nu}_V\,,
\end{equation}
where $\epsilon$ is consequentially a rather small dimensionless quantity as it is loop-induced. Explicitly, one finds that 
\begin{equation}
\epsilon =\frac{g_D e}{12\pi^2} \sum_i ~(\eta_i N_{c_i}Q_{em_i}Q_{D_i})~ ln \frac{m^2_i}{\mu^2}\,,
\end{equation}
with $g_D$ being the $U(1)_D$ gauge coupling (so that $\alpha_D=g_D^2/4\pi$)  and $m_i(Q_{em_i},Q_{D_i}, N_{c_i})$ are the mass (electric charge, dark charge, number of colors) of the 
$i^{th}$ PM field. Here, $\eta_i=1(1/4)$ if the PM is a vector-like fermion (complex scalar). In a fully UV-complete theory, as is the eventual goal of the types of setups which we that will be 
discussing below (although this spect will not be employed explicitly), we might expect this sum to satisfy the condition that 
\begin{equation}
\sum_i ~(\eta_i N_{c_i} Q_{em_i}Q_{D_i})=0\,,
\end{equation}
so that $\epsilon$ will be both finite and, in principle, calculable, due to cancellations taking place between the various contributions arising from group theoretical requirements.  For a set of PM fields 
of similar mass, we might be expect $\epsilon$ to lie in the range of $\sim 10^{-(3-4)}$ which is roughly that required to satisfy both experimental search constraints and to obtain the observed DM 
abundance for sub-GeV DM and DP masses.   

As has been discussed in earlier work\cite{Rizzo:2018vlb,Rueter:2019wdf,Rueter:2020qhf,Wojcik:2020wgm,Rizzo:2021lob,Rizzo:2022qan,Rizzo:2022jti,Rizzo:2022lpm,Rizzo:2023qbj,Rizzo:2023kvy,Rizzo:2023djp}, setups such 
as these are subject to numerous constraints both from measurements at accelerators at high energy as well as those from DM direct searches at low scales.  
It is to be noted that within this class of sub-GeV DM models, both cosmology and astrophysics also place very tight bounds on the velocity-weighted cross sections for the DM annihilation into 
(charged) SM particles as these must be significantly suppressed at later times (relative to the typical thermal annihilation target $\sigma v_{rel}\sim 3\times 10^{-26}~$cm$^3$ sec$^{-1}$ required 
at freeze-out\cite{Steigman:2015hda,Saikawa:2020swg}), \ie, during the CMB\cite{Planck:2018vyg,Slatyer:2015jla,Liu:2016cnk,Leane:2018kjk} and at 
present\cite{Koechler:2023ual,DelaTorreLuque:2023cef}. 
This implies that such cross sections must display a significant temperature ($T$) dependence and so {\it cannot} correspond to an $s$-wave annihilation process. Instead such processes must 
occur, \ie,  ($i$) via $p-$wave annihilation so that there is a $v^2 \sim T$ suppression at later times, as may be realized in the case of complex scalar or Majorana fermion 
DM{\footnote {However\cite{Belanger:2024bro}, in 
the case of DP exchange one must also take care as this $p-$wave-induced $v^2$ suppression alone might not be {\it sufficiently suppressed} at later times in certain parameter space regimes 
if the observed relic density is achieved via a resonant enhancement.}}. Another possibility is that the observed relic density may be achieved $(ii)$ through the co-annihilation mechanism,  as can be 
realized in the case of pseudo-Dirac DM with a sizable mass splitting, so that annihilation rates are exponentially Boltzmann-suppressed at later times again due to the far lower 
temperatures\cite{Brahma:2023psr,Balan:2024cmq,Garcia:2024uwf,Mohlabeng:2024itu}. Both of these cases, $(i)$ and $(ii)$, will be encountered simultaneously in our discussion below.

Returning now to the KM mechanism itself, one may wonder how the newly required PM particles will fit together into a single picture with those of the SM as well as with the DM itself and this has 
been a major effort in part of our recent work\cite{Rizzo:2018vlb,Rueter:2019wdf,Rueter:2020qhf,Wojcik:2020wgm,Rizzo:2021lob,Rizzo:2022qan,Rizzo:2022jti,Rizzo:2022lpm,Rizzo:2023qbj,Rizzo:2023kvy,Rizzo:2023djp}, 
with the eventual goal being the construction of an at least partially UV-complete setup and at which we've been only at best partially successful. An important hint to  
this process is the realization that the abelian $U(1)_D$ gauge group is itself likely to be the low energy remnant of some larger non-abelian group, $G_D$, potentially also simultaneously involving 
at least some of 
the PM, DM and SM fields and which is broken at a large (> a few TeV) scale, the same scale at which the PM obtains its mass. It has been argued in earlier work, as well as by other sets of authors\cite{Rizzo:2018vlb,Rueter:2019wdf,Kim:2019oyh,Rueter:2020qhf,Wojcik:2020wgm,Rizzo:2021lob,Rizzo:2022qan,Wojcik:2022rtk,Rizzo:2022jti,Rizzo:2022lpm,Wojcik:2022woa,Carvunis:2022yur,Verma:2022nyd,Rizzo:2023qbj,Wojcik:2023ggt,Rizzo:2023kvy,Rizzo:2023djp}, that over a significant range of low energy couplings, the RGE running of $\alpha_D$ indicates that this new scale 
is possibly not very far away due to the eventual loss of perturbativity. This perhaps takes place at a scale  as low as the $\sim 10$ TeV mass 
range\cite{Davoudiasl:2015hxa,Reilly:2023frg,Rizzo:2022qan,Rizzo:2022lpm}, depending, of course, upon the details of the low energy field content. The basic idea is that the embedding of 
$U(1)_D$ into this larger non-abelian group, $G_D$, which is assumed to be asymptotically free, reverses the `bad' RGE running {\it before} this high scale is reached thus avoiding this problem. 

In the simplest setup considered in earlier work, it has been assumed that $G_D=SU(2)_I\times U(1)_{Y_I}$\cite{Bauer:2022nwt}, into which $U(1)_D$ can be embedded in a quite 
familiar fashion, and wherein at 
least some of the PM fields lie in doublet representations along with SM fields with which they share common strong and electroweak properties{\footnote {This was motivated by our study by 
earlier work on $E_6$-type gauge models\cite{Hewett:1988xc}}}. One may then imagine some further completion wherein a larger `unification' group, $G_U$, breaking down at an even higher scale to, \eg,  $G_D\times G_{SM}$ or an even some larger group of which $G_D\times G_{SM}$ is a 
subgroup. In general setups of such types, one might further anticipate  
that the DM field is also found to be a component of some non-trivial representation (\ie, a non-singlet) of $G_D$ along with other SM singlet fields that have, \eg,  $Q_D=0$.  
However, in our recent studies, the DM was generally treated as a complex $SU(2)_I$ (or, more generally, $G_D$) singlet for purposes of simplicity; this is certainly not the general situation that 
we might expect to encounter in a more UV-complete model; the corresponding implications of an even trivially extended sector containing the DM within the PM framework remains almost 
completely unexplored. What happens when we make this rather simple change to the basic PM setup? 
The purpose of the present paper is to address this question, remedying this situation by constructing and examining the consequences of a rather simple toy model where it is no longer the case that 
the DM state sits alone in the dark sector, \ie, where the DM, here assumed to be a vector-like fermion (VLF), lies in an $SU(2)_I$ doublet together with another SM singlet having $Q_D=0$. Even 
this rather simple setup will be shown to lead to some interesting new predictions and phenomenology.

The outline of this paper is as follows:  Following this Introduction, in Section 2,  we present the basic structure of our toy model, wherein the VLF DM sits in an $SU(2)_I$ doublet, to set up 
the phenomenological analyses that follows.  The mass patterns of the DM field, $\chi$, and it's $SU(2)_I$ partner with $Q_D=0$, $\psi$, are then examined. The phenomenological need to make 
this fermionic DM a Majorana or pseudo-Dirac state due to late times annihilation constraints necessitates the introduction of an $SU(2)_I$ triplet Higgs field $T$, having a small vev, $v_T\sim 1$ GeV, 
which also sources the DP mass. This implies the existence of additional (and as we'll see heavy) dark Higgs degrees of freedom that are not usually encountered in PM models, having 
interesting FCC-hh collider signatures. As expected, the originally Dirac field $\chi$ splits into two distinct mass eigenstates, $\chi_{1,2}$ with the lighter state corresponding to the DM, which 
are admixtures of the both the left-handed and right-handed components of $\chi$ and their conjugates. This 
mixing angle, $x=\sin^2 \theta$, controls the relative admixture of the Majorana and pseudo-Dirac nature of the DM and plays an important role in controlling its coupling to the DP. 
The possible mixing of the $Q_D=0$, SM-singlet $\psi$ with the SM neutrino via the usual $SU(2)_I$-breaking doublet, $H_D$, which has a large vev, $v_D\gsim 10$ TeV, is then 
analyzed in two toy setups (with the second one being slightly more realistic) where it is found that this mixing not only boosts the $\psi$ mass up to the scale of the $H_D$ vev but leads to a 
corresponding suppression in the anticipated SM Dirac neutrino mass (in comparison to, say, the electron mass) by a factor of roughly $\sim 10^5$.  

In Section 3, we discuss the calculations of the DM relic density as well as its elastic scattering with electrons in this setup via the exchange of a DP.  In particular for the relic density calculation, these 
interaction rates are shown to be not only controlled by the $\chi_{1,2}$ mass splitting and the ratio of the DP and DM masses but are also quite sensitive to the value of the mixing parameter, $x$.  
Section 4 contains a general discussion of the masses and mixings amongst the new dark Higgs doublet and triplet fields required within the $G_D=SU(2)_I\times U(1)_{Y_I}$ framework under 
the assumption of CP-conservation. In Section 5, in order to make the phenomenological study of the dark Higgs states (as well as $\psi$) at colliders more tractable by reducing the number of model 
parameters, we introduce a toy benchmark model which allows the relative spectrum of all of the new heavy states, including the $SU(2)_I$ gauge bosons $W_I$ and $Z_I$, as well as their various 
mixings, to be calculable up to an overall mass scale set by the $H_D$'s vev, $v_D$.  Within this framework, we first show that the existing LHC search limit on the $Z_I$ pushes $v_D$ above 
$\simeq 13.5$ TeV so that the production and study of these new Higgs states will require the 100 TeV FCC-hh to obtain a sufficiently large cross section and data sample for further analysis. 
The phenomenology of these new scalars as well as that of $\psi$ is then examined. Finally, a summary and our conclusions can be found in Section 6.


\section{Background, Mass Mixing and General Setup}

As per the above discussion, we will assume for concreteness in the following analysis that $G_D=SU(2)_I \times U(1)_{Y_I}$ acts as the dark sector/PM gauge group as was done in previous 
works\cite{Rizzo:2018vlb,Rueter:2019wdf,Rueter:2020qhf,Wojcik:2020wgm,Rizzo:2021lob,Rizzo:2022qan,Rizzo:2022jti,Rizzo:2022lpm,Rizzo:2023qbj,Rizzo:2023kvy,Rizzo:2023djp}. Furthermore, 
as was also done previously, we will assume that $G_D$ is broken down to the familiar $U(1)_D$ associated with the DP, $V$, at a high mass scale in the $\sim$ 10's 
of TeV range via the vev, $v_D$, of an $SU(2)_I$ dark Higgs isodoublet, $H_D$, in analogy with the SM. Recall here that, again similar to the SM case, the dark charge is given by the combination 
$Q_D=T_{3I}+Y_I/2$.  At this large scale, both the PM fields and the addition heavy gauge bosons, $W_I,Z_I$, in $G_D$ will obtain their masses. 

In generalizing from our previous treatments of SM singlet DM, which is also an $SU(2)_I$ isosinglet, we will consider the most obvious and next simplest non-trivial possibility. Here we analyze a 
setup wherein the $Q_D=1$ DM, $\chi$, which will assume to be a VLF to avoid gauge anomalies,\etc,  will share an $SU(2)_I$ doublet, $X$, with another $Q_D=0$, SM singlet VLF, $\psi$, \ie, 
\begin{equation}
X_{L,R}=\begin{pmatrix}\chi \\ \psi \\ \end{pmatrix}_{L,R} \,,
\end{equation}
for which we can write a general mass term in the Lagrangian as
\begin{equation}
{\cal L}_{Dark} = -m_D\bar X_L X_R -\frac{1}{2}(y_L \bar X_L^cX_L+y_R \bar X_R^cX_R)T+{\rm {h.c.}}\,,  
\end{equation}
where $m_D$ is a gauge invariant VLF Dirac mass for $X$ and $T$ is an $SU(2)_I$ triplet Higgs field whose uppermost component obtains a small [since we observe that it has $Q_D(v_T)=2$],  
$Q_D$-violating vev,  $<T>=(v_T,0,0)^T/\sqrt 2$,  generating distinct Majorana mass terms, $\mu_{L,R}=y_{L,R}v_T/\sqrt 2$, for the $\chi_{L,R}$ fields via their associated Yukawa couplings which 
we might expect to be $O(1)$. As the DM mass will be required in our analysis to be in the sub-GeV range, $m_D$ will also assumed to be $\sim 100$ MeV. 
Since this same vev, $v_T$, will simultaneously also generate the needed mass term for the DP, \ie, $m_V=2g_Dv_T$, we expect that $v_T\sim 100-1000$ MeV also. For later convenience and 
closely following, \eg, Ref. \cite{Abdullahi:2023tyk}, we can define two combinations of these Majorana masses: $\mu =(\mu_R+\mu_L)/2$ and $\Delta=(\mu_R-\mu_L)/2$.  
Here we stress that $m_D$ and $\mu_{L,R}$ are all expected to be of somewhat comparable magnitude, $\sim m_V$, but with the inequality $m_D>\mu$ assumed to be satisfied. While the field 
$\psi$ remains a simple massive Dirac fermion at this point in the discussion, a $2\times 2$ mass matrix, $M_\chi$, is generated between the chiral components of $\chi$, which, again following 
Ref.\cite{Abdullahi:2023tyk}, arises from the coupling structure
\begin{equation}
-\frac{1}{2}\begin{pmatrix} \bar {\chi^c}_L, & \bar \chi_R \\ \end{pmatrix}  \begin{pmatrix} \mu_L & m_D \\ m_D & \mu_R \\  \end{pmatrix} \begin{pmatrix}\chi_L \\ \chi^c_R \\ \end{pmatrix}+h.c.\,,
\end{equation}
which can be diagonalized through a complex orthogonal mixing by defining the states $\chi_{1,2}$ as  
\begin{equation}
\chi_L=ic_\theta \chi_{1L}+s_\theta \chi_{2L},~~\chi_R=is_\theta \chi_{1R}^c+c_\theta \chi_{2R}^c \,,
\end{equation}
where one finds that 
\begin{equation}
\tan 2\theta=\frac{m_D}{2\Delta}\,,
\end{equation}
which subsequently leads to the mass eigenvalues $m_{1,2}=m_D\mp \mu$ so that we can define $m_2=m_1(1+\delta)$. $\chi_1$, which we recall carries $Q_D=1$, is now the lightest dark state 
and will be identified with DM which we assume still satisfies $m_1 \sim m_V \sim 0.1-1$ GeV as noted above.  In terms of these physical $\chi_{1,2}$ mass eigenstates and the mixing 
angle $\theta$, their interactions of the DP can be now expressed as 
\begin{equation}
{\cal L}_{int}=g_DJ_\mu^DV^\mu=\frac{g_D}{2}\Big[\cos 2\theta \big(\bar\chi_1\gamma_\mu \gamma_5 \chi_1 -(1\to 2)\big)+i\sin 2\theta \big(\bar\chi_2\gamma_\mu \chi_1 -(1\leftrightarrow 2)\big)\Big]V^\mu\,.
\end{equation}
From this we can see that the weighting of the purely Majorana vs. pseudo-Dirac annihilation channels at the `parton' level before thermal averaging is quite sensitive to the mixing angle as well as 
the mass splitting $\delta$ and this will have important implications for the discussion in the next Section. A qualitatively similar coupling structure in the $\chi_{1,2}$ mass eigenstate basis will also 
be generated with the physical light dark Higgs, $h_D$, which remains in the spectrum after symmetry breaking. However, as we'll discuss further below, it will play no vital role in the relic density 
determination even though the relevant couplings, $\sim \mu/v_T$, might expected to be sizable, \ie, $O(1)$.

In principle, the on-shell decay $\chi_2 \to \chi_1 V$ is an allowed mode while here being suppressed by $\sin^2 2\theta$. However, this requires that $r\delta>1$, where $r=m_1/m_V$, which will 
lie outside of the range of parameters that we will be concerned with below; this is especially so since we will require $r<1$ to avoid the $s$-wave $\chi_1 \chi_1 \to 2V$ reaction from 
occurring{\footnote {Stronger constraints on $r$ as a function of $\delta,\epsilon,\alpha_D$ may apply\cite{Rizzo:2020jsm} when we consider the possibility of additional dark initial state 
radiation emitted in the, \eg, $\chi_1\chi_1\to e^+e^-$ process.}}.  Instead, the 
3-body process $\chi_2 \to \chi_1 V^*, V^*\to e^+e^-$ will be the more relevant one leading to a suppressed partial width given roughly by\cite{Garcia:2024uwf} 
\begin{equation}
\Gamma_{3-body}\simeq \frac{4m_2^5}{15\pi m_V^4}\alpha_D \alpha \epsilon^2 \delta^5 \sin^2 2\theta\,,  
\end{equation}
when the electron mass has be neglected, so that $\chi_2$ is very likely to be rather long-lived. 

Let us now return to the SM singlet, $Q_D=0$, state $\psi$ and re-introduce another set of familiar fields in addition to $X$ above: a generic SM lepton left-handed isodoublet, $L=(\nu,e)^T_L$, a 
SM isosinglet right-handed neutrino, $\nu_R$, the usual SM Higgs isodoublet field, $H$, as well as the dark Higgs $SU(2)_I$ doublet mentioned above, $H_D$, which is also a SM singlet. As 
previously noted, all of the SM fields have $Q_D=0$. Then we can write down several mass terms which will involve only these electrically neutral, $Q_D=0$ fields: 
\begin{equation}
{\cal L}_{Mass} = m_D\bar \psi_L \psi_R +y_\nu \Bar L_L \nu_RH+y\Bar \psi_L\nu_R H_D +{\rm h.c.}\,,  
\end{equation}
which leads to the corresponding to the $2\times 2$ matrix ${\cal M}$ in the $(\nu,\psi)_{L,R}$ basis:
\begin{equation}
{\cal M}= \begin{pmatrix} m_{SM} & 0 \\ M_I & m_D \\  \end{pmatrix}\,,
\end{equation}
where the would-be Dirac SM neutrino mass is defined to be $m_{SM}=y_\nu v_{SM}/\sqrt {2}$, and one that we might very naively expect to be of order the electron mass, $M_I=yv_D/\sqrt {2}$ is 
a mass of order that of the PM fields and the heavy gauge bosons associated with the $SU(2)_I$ breaking scale, $\sim 10$ TeV, and $m_D\sim 0.1$ GeV is the DM Dirac mass as defined above. 
The matrix ${\cal M}$ can be diagonalized by the familiar bi-unitary transformation matrices, $U_{L,R}$, acting separately on the left- and right-handed fermion fields (which in this case can both 
be chosen to be simple real orthogonal matrices)
\begin{equation}
M_{diag}=U_L {\cal M} U_R^\dagger\,,  
\end{equation}
with, as usual, described by the corresponding mixing angles, $\theta_{L,R}$, and with mass eigenvalues that can then determined by use of the standard relationships: 
\begin{equation}
M_{diag}^2=M_{diag}^\dagger M_{diag}=M_{diag}M_{daig}^\dagger=U_L {\cal M}{\cal M}^\dagger U_L^\dagger =U_R {\cal M}^\dagger{\cal M} U_R^\dagger\,.  
\end{equation}
From these expressions we find that both of these mixing angles are quite tiny due to the large mass hierarchies, \ie,  
\begin{equation}
\theta_L \simeq -m_{SM}/M_I \sim -10^{-7}, ~~~\theta_R\simeq m_D/M_I \sim 10^{-5}\,,  
\end{equation}
with the corresponding the (Dirac) see-saw mass eigenvalues being given by
\begin{equation}
m_{-} \simeq m_{SM}\frac{m_D}{M_I} \sim 10^{-5}m_e \sim{\rm eV's},~~~ m_{+}\simeq M_I\sim 10 ~{\rm TeV}\,.  
\end{equation}
Here we have assumed that $m_D>$ several - 10 MeV so that the DM mass also satisfies the current low DM mass nucleosynthesis bounds as discussed in Refs.\cite{Giovanetti:2021izc,Chu:2022xuh,Sabti:2021reh,Sabti:2019mhn} and thus one has (for this single generation toy 
model) that $m_{SM}^2<<m_D^2$. While the anticipated Dirac $\nu$ mass is reasonably suppressed by $\sim 10^5$ compared to `naive expectations' and goes quite far in the right direction, 
it is unfortunately still remains a factor of roughly $\sim 10^2$ or so too large in comparison to what we might desire based on the values actually realized in measurements from oscillation 
experiments. Still, an interesting step in the right direction.

An essentially identical result is obtained even when one tries to add some more realism to this toy model setup by including, \eg, actual VLF PM fields into this mix. A simple, straightforward 
and familiar example of this, which we've encountered in earlier works and will encounter again below, \eg, \cite{Rueter:2019wdf}, is to add the lepton-like, SM isodoublet VLF fermion fields 
with $Q_D=1$ in the following manner [where the $SU(2)_L(SU(2)_I)$ group in this specific example acts vertically(horizontally)]:
\begin{equation}
{\cal B}_L=\begin{pmatrix} N & \nu \\ E & e \\  \end{pmatrix}_L,~~ {\cal N}_R=\begin{pmatrix}N \\ E\\ \end{pmatrix}_R \,,
\end{equation}
with the SM $(\nu,e)_L^T$ and  PM $(N,E)_L^T$ fields together now forming a chiral fermion bi-doublet under the $SU(2)_L\times SU(2)_I$ product group with $Y_I(Y_{SM})=1(-1)$, while the 
right-handed 
PM remains an $SU(2)_I$ singlet. Numerous additional mass terms can now be generated, in principle, possibly requiring an extension (or a `translation') of the Higgs sector already introduced 
above, but now in a slightly more generalized context as the SM lepton doublet is here embedded into the fermion bi-doublet.   A subset of these possible new terms that will allow for the DM 
to maintain its stability can be generically summarized as 
\begin{equation}
{\cal L'}_{Mass}=y_1 \bar {\cal B}_L {\cal N}_R H_1+y_2 \bar {\cal B}_L \nu_R H_2 +y_3 \bar X_L{\cal N}_R H_3+{\rm h.c.}\,,
\end{equation}
where we need to identify and understand the nature of the three Higgs fields, $H_{1-3}$, two of which we've actually (at least partially) already encountered above in an another guise. First, we 
see that $H_1$ is just the SM singlet, $SU(2)_I$ doublet Higgs field, \ie, $H_D$, introduced above to break $SU(2)_I\times U(1)_{Y_I}\to U(1)_D$ generating the $W_I$ and $Z_I$ masses as well as 
that for the lepton-like PM fields here in the term proportional to $y_1$; we'll refer to these Dirac $N,E$ masses here simply as $m_N$. Recall that since $H_D$ is a SM singlet, both of its components 
are therefore neutral having $Q_D=0$ or 1.  Thus, in principle, both of these components may obtain vevs and it is the $Q_D=0$ one, which we'll designate for this discussion as $v_2$ but more 
generally $v_D$, which obtains the large $\gsim 10$ TeV scale vev in this case to generate $m_N$. The 
other component also obtains a smaller vev, $v_1$, breaking $Q_D$, thus contributing to the DP mass, and so it is required to be much smaller, $\sim 0.1-1$ GeV. This vev couples the states 
$\nu_L-N_R$ as well as $e_L-E_R$ so that, \eg, the $Q_D$-violating (and far dominant) decay paths 
$E(N)\to e(\nu) V$ can occur. The term proportional to $y_2H_2$, on the other-hand, is also mandatory as it generates both the SM electron mass as well as the Dirac $\nu$ mass term employed 
above. Thus, we see that the familiar SM Higgs with the vev $v_{SM}$, is actually just the $Q_D=0$ half of $H_2$ ,which is an $SU(2)_l\times SU(2)_L$ bi-doublet. The `other' $SU(2)_L$ doublet 
with $Q_D=1$ which resides in $H_2$ also has a neutral component which can obtain a small vev, $v' \lsim 0.1-1$ GeV as it similarly breaks $U(1)_D$ in addition to $SU(2)_L$. This vev also 
allows the decays $E(N)\to e(\nu) V$ to occur but with the opposite chiral structure than that produced by $v_1$. Finally, $H_3$ is {\it also} seen to be a bi-doublet, but is distinct from $H_2$  
with a different hypercharge and so having only a single $Q_D=1$ vev, $\tilde v\lsim 0.1-1$ GeV, in order to avoid the possible kinematically allowed decay $\chi_i \to \nu V$ from occurring.  

Denoting $m_N=y_1v_2/\sqrt 2$, $A=y_1v_1/\sqrt 2$, $B=y_3\tilde v/\sqrt 2$ and $C=y_2v'/\sqrt 2$, in the $(\nu,\psi,N)^T_{L,R}$ basis, we now obtain a $3\times 3$ extended version 
of the matrix ${\cal M}$ above as
\begin{equation}
{\cal M}_3=\begin{pmatrix} m_{SM} & 0 & A\\ M_I& m_D & 0\\ C & B & m_N\\ \end{pmatrix}\,,
\end{equation} 
where $A\sim B\sim C\lsim 0.1-1$ GeV but $m_N\sim $ a few TeV or more if we assume that the $y_{1-3}\sim O(1)$. Diagonalizing this matrix leads to the three eigenvalues $\simeq m_N,M_I$ and 
$\simeq m_{SM}m_D/M_I$, indeed replicating our previous result, and something that should perhaps have been expected based on the locations of the zero entries appearing in ${\cal M}_3$. 
This limited extension to the original toy model, in particular, has unfortunately failed to provide us with the added flexibility required to further increase the needed suppression of the SM neutrino 
mass. On top of this, it is clear that since all of the off-diagonal elements are quite small $< 1$ GeV, the mixing between the three mass eigenstates is found to be very highly suppressed as was 
the case in the previous $2\times 2$ example.

\section {DM Relic Density and DM-Electron Scattering}

Let us now return to the DM sector.  
Given the Lagrangian above describing the $\chi_i$ interactions with the DP, we see that we need to consider both the direct annihilation processes, \ie, the $\bar\chi_1\chi_1, \bar \chi_2\chi_2\to$ 
SM channels as well as the co-annihilation $\bar\chi_1\chi_2+ \rm{h.c.}\to$ SM process in order to determine the DM relic density. During the CMB, as well as at present times, while the 
$\bar\chi_1\chi_1$ annihilation process will be $v^2$ suppressed (as it is a $p-$wave due to the presence of the $\gamma_5$ in the diagonal coupling), the co-annihilation channel reaction is 
instead an $s-$wave but one which will 
become significantly Boltzmann-suppressed at lower temperatures due to the assumed sizable $\chi_2-\chi_1$ mass splitting, $\delta$. The corresponding $ \bar \chi_2\chi_2$ initiated process is 
then found to be {\it both} $p$-wave as well as doubly Boltzmann-suppressed in later epochs.

In principle, a dark Higgs exchange, which can be seen to arise due the generation of the Majorana mass terms above, can also yield potentially significant $p-$wave contributions to both the 
$\bar\chi_1\chi_1, \bar \chi_2\chi_2\to h_D^* \to$ SM annihilation processes. This occurs via an $s$-channel exchange when one accounts for the allowed mixing of the dark Higgs with the familiar 
SM one. Furthermore 
unitarity arguments\cite{Li:2024wqj} suggest that this dark sector Higgs state is not very heavy in comparison to the DP, $m_{h_D}<m_V/(8\alpha_D)^{1/2}$, so that it's contribution will not 
automatically be suppressed due to its relatively large mass. However, as noted, in order to couple to the SM fields in the final 
state, $h_D$ must first mix with the SM Higgs, via a mixing angle, $\theta_H$, that we already now know must be substantially suppressed to avoid a sizable $H(125)\to$ invisible  
branching fraction, $B_{inv}\lsim 0.1$\cite{ATLAS:2023tkt,CMS:2023sdw}. In fact, one finds that to satisfy this constraint we must have $|\theta_H|\lsim 2\times 10^{-4}$. Further, since only the light 
fermions, such as $e,\mu$, are likely to 
be kinematically accessible as on-shell SM final states, the relevant Yukawa coupling of the SM Higgs to these states will itself already be suppressed (at least) by a factor of 
$\sim m_\mu/v_{SM} \simeq 4\times 10^{-4}\sim \epsilon$. Thus, due to this double suppression, we can safely ignore any possible $h_D$ exchange effects in the discussion that follows. Interestingly, 
in the situation when the DP dominantly decays to on-shell pairs of electrons or muons (as would be the case if $2m_1>m_V$) then additional constraints arising from the recent ATLAS and 
CMS null searches\cite{ATLAS:2024zxk,CMS:2024jyb} for pairs of lepton-jets arising from the decay $H(125)\to 2V \to 4$ leptons would also be applicable. In such a case, these limits would strongly 
suggest that this particular mass ordering cannot be realized unless the mixing between the SM Higgs and $h_D$ is {\it very} highly 
constrained by at least one or two additional orders of magnitude, depending upon the DP mass, beyond that just described.

Returning now to the calculation of the relic density, we can symbolically express the overall relative velocity-weighted annihilation DM cross section in the form
\begin{equation}
\sigma v={\cal F}^2 \frac{\tilde N}{(1+F)^2}\Big[ (1-W)\big(\sigma_{11}+F^2\sigma_{22}\big)+WF\sigma_{12}\Big]\,,
\end{equation}
where $W=4\sin^2\theta \cos^2 \theta=4x(1-x)$, $\tilde N$ is an overall numerical factor $\simeq 8.52 \sigma_0$, with $\sigma_0=10^{-26}~$cm$^3$ sec$^{-1}$ being the general scale of the  
required DM annihilation cross section to achieve the observed relic density, and the $\sigma_{ij}$ are relative velocity-weighted effective reduced cross sections, corresponding to the 
different annihilation channels, which are functions of the parameters $\delta$, $r=m_1/m_V$ {\footnote {Remember that to avoid possible $s$-wave annihilation into the $2V$ final state, $r$ must 
be bounded from above as will be discussed further below.}}, as well as the scaled total decay 
width of the DP, $G_V=\Gamma_V/m_V$. Note that since the DP's SM decay partial widths are highly suppressed by $\epsilon^2 \sim 10^{-7}$, $\Gamma_V/m_V$ is obviously quite 
numerically sensitive to the opening up of thresholds for the various $V\to \chi_i\chi_j$ on-shell decay channels which are shown by the solid curves in Fig.\ref{fig0}. Also, as noted above, the recent 
ATLAS and CMS null searches\cite{ATLAS:2024zxk,CMS:2024jyb} for collimated pairs of lepton-jets arising from the decay $H(125)\to 2V \to 4$ leptons, strongly suggests that $r<1/2$, unless 
the mixing of $h_D$ with the SM Higgs is extremely small, so that it is {\it likely} that the DP can directly decay invisibly to DM.

The $\sigma_{ij}$ cross section are then given by the simple expressions (in the limit that all final state masses can be neglected, \eg, for the $e^+e^-$ final state)
\begin{eqnarray}
&~~~~&\sigma_{11}=\frac{1}{3} P\tilde s \beta_1^2\nonumber\\
&~~~~&\sigma_{22}=\frac{1}{3} P\tilde s \beta_2^2\nonumber\\
&~~~~&\sigma_{12}=2P \Big(e_1e_2+r^2(1+\delta)-\frac{1}{3}\tilde p^2\Big)\,,
\end{eqnarray} 
where 
\begin{equation}
\tilde p^2=\frac{\big[\tilde s^2-2\tilde s (1+(1+\delta^2))r^2+(2\delta+\delta^2)^2\big]}{4\tilde s},~~~~ e_{1,2}=\frac{\tilde s\mp (2\delta+\delta^2)}{2\sqrt{\tilde s}}\,,
\end{equation}
and $P^{-1}=(\tilde s-1)^2+G_V^2$, with $\tilde s=s/m_V^2$ ,and where $\beta_i^2=1-4m_i^2/s$, respectively. In addition, one has that 
\begin{equation}
F=(1+\delta)^{3/2}e^{-\delta x_F}\,,
\end{equation}
with, as usual, $x_F=m_1/T_F\simeq 20$, $T_F$ being the thermal freeze-out temperature. Employing the value of $\tilde N$ above, one then finds that 
\begin{equation}
{\cal F} =\Big(\frac {g_D\epsilon}{10^{-4}}\Big)~ \Big(\frac{100 ~{\rm MeV}}{m_V}\Big)\,,
\end{equation}
is just an {\it a priori} unknown model-dependent pre-factor, scaled to typical parameter values. The partial width for the color-singlet decays of the form $V\to \bar f_1f_2+h.c.$, $\Gamma_{12}$,  
described by a general coupling structure $\sim \gamma_\mu (v-a\gamma_5)\epsilon_V^\mu$,  is given by the somewhat familiar expression 
\begin{equation}
\Gamma_{12}=\frac{[(m_V^2+m_1^2-m_2^2)^2-4m_V^2m_1^2]^{1/2}}{3m_V^5}\Big( [2m_V^4-m_V^2(m_1^2+m_2^2)-(m_2^2-m_1^2)^2](v^2+a^2)+6m_1m_2m_V^2(v^2-a^2)\Big)\,,
\end{equation}
which can be evaluated for the various relevant final states (including the leptonic SM modes). These individual contributions can be then bee appropriately summed over to obtain the scaled 
total DP decay width, $G_V$, depending upon where one lies in the model phase space; this expression simplifies significantly in the case of diagonal couplings. 

The required value of the overall scale factor ${\cal F}$ can then be determined by demanding that the total, now {\it thermally averaged} annihilation cross section, 
satisfy the equality $\overline{\sigma v}\simeq 2.6(4.4)\sigma_0$, which for self-annihilating Majorana (co-annihilating) DM with a mass lying in the range $\sim 10-200$ MeV, is roughly the  
value needed\cite{Steigman:2015hda,Saikawa:2020swg} to obtain the DM relic density observed by Planck, as a function of the specific model parameters $r,\delta$ and $x=\sin^2 \theta$. To 
perform this calculation we must also sum over the various SM final states, $e^+e^-,\mu^+ \mu^-$, \etc, which are kinematically accessible. Here, for both simplicity and numerical purposes, we 
will consider {\it only} the $e^+e^-$ mode but it is important to recall the possible likely contributions of these other SM final states depending upon the $\chi_{1,2}$ masses and so one will need to 
approximately rescale our results below by a factor of $[{\cal R}=\overline{\sigma v}(\to {\rm all})/\overline{\sigma v}(e^+e^-)]^{-1/2}$. 

Given a set of input parameters, \ie, $x=\sin^2\theta ,r$ and $\delta$, we can integrate the above cross section expressions, weighted by the appropriate relativistic Fermi-Dirac thermal 
distributions, to obtain the annihilation cross section and then ask for that value of the overall numerical factor ${\cal F}$ which is necessary to obtain the observed relic density. In considering the 
results of this 
calculation, we may be worried about avoiding the kinematic regions that allow for the on-shell $\chi_i\chi_j \to 2V$ processes as these are $s$-wave and so can make potentially 
dangerous contributions to DM annihilation during the CMB and later epochs as noted above. These kinematic boundaries for the case of $T=0$ are shown as the dashed curves in Fig.~\ref{fig0}. 
However, due to the strong Boltzmann suppression experienced at the lower temperatures at the time of the CMB only the constraint $r<1$ arising from the $\chi_1\chi_1 \to 2V$ process is 
actually relevant in practice.

\begin{figure}[htbp]
\centerline{\includegraphics[width=5.0in,angle=0]{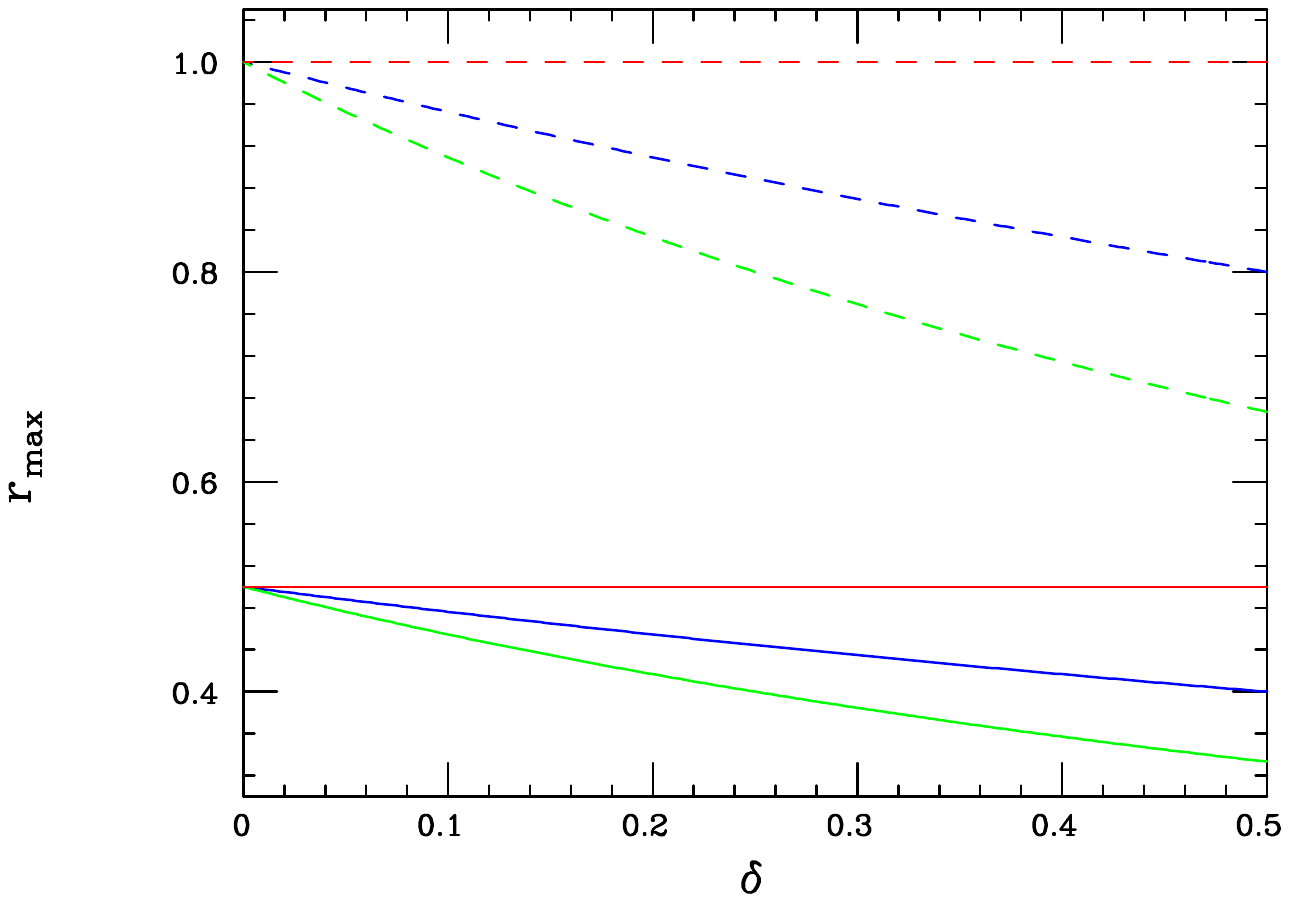}}
\vspace*{-1.3cm}
\caption{The maximum value of $r$, \ie, $r_{max}$, as a function of $\delta$ corresponding to various kinematic thresholds. The red (blue, green) dashed curves at the top of 
the Figure correspond to the smallest values of $r$ (when $T=0$) at which the on-shell processes $\chi_1\chi_1(\chi_1\chi_2, \chi_2\chi_2) \to 2V$ become kinematically allowed as discussed in 
the text. The red (blue, green) solid curves at the bottom of the Figure correspond to the values of $r$ below which the DP decays $V\to \chi_1\chi_1(\chi_1\chi_2,\chi_2\chi_2)$, can occur 
on-shell and so significantly increasing the DP's total width as is also discussed in the text.}
\label{fig0}
\end{figure}

Figs.~\ref{fig1} and Fig.~\ref{fig2} show the values of the scale factor ${\cal F}$ as a function of $r$ for different choices of the values of $\delta$ and $x$ that are necessary to obtain the observed DM 
relic density in the current setup. Overall, values of ${\cal F}$ roughly within an order of magnitude of unity are observed to be required over the entire parameter range of interest and clearly a 
significant parameter range is allowed. 
In all cases, the overall qualitative common nature of these results is easily understood as due to the cross section enhancement arising from the existence of the 
DP resonance (appearing in the vicinity of $r\lsim 1/2$), thus leading to a suppression of the necessary value of ${\cal F}$ in that parameter space region which is common to all cases. At the extreme 
values of $x$ only one of the three sub-processes contributes and we of course then return to the purely Majorana case ($x=0$) or to the purely pseudo-Dirac case ($x=0.5$). For values of $x$ away 
these extremes we observe an admixture of contributions causing complex behaviours near the resonance region as the various cross sections `see' this resonance at slightly different values of $r$ 
(but always below $r=0.5$) due to their differing thermally weighted velocity dependencies. Two general features that we observe are that the curves both deepen (by roughly a factor of 3) going to 
smaller values of ${\cal F}$ and become broader, \eg, by roughly a factor of 2 at ${\cal F}$ values near unity, as $x$ increases.  
Of course, when $x=0$ there is little $\delta$ sensitivity but this becomes quite significant as one moves towards $x=0.5$, as might be expected since then only the co-annihilation process is 
relevant in such a limit. Again, as might be expected, as $\delta$ increases, the minimum of the curve for ${\cal F}$ moves to smaller values of $r$, as the $\chi_2$ mass (relative to that of $V$) 
is increasing, and moves to larger values due to the strengthening of the Boltzmann suppression.

\begin{figure}[htbp]
\centerline{\includegraphics[width=4.2in,angle=0]{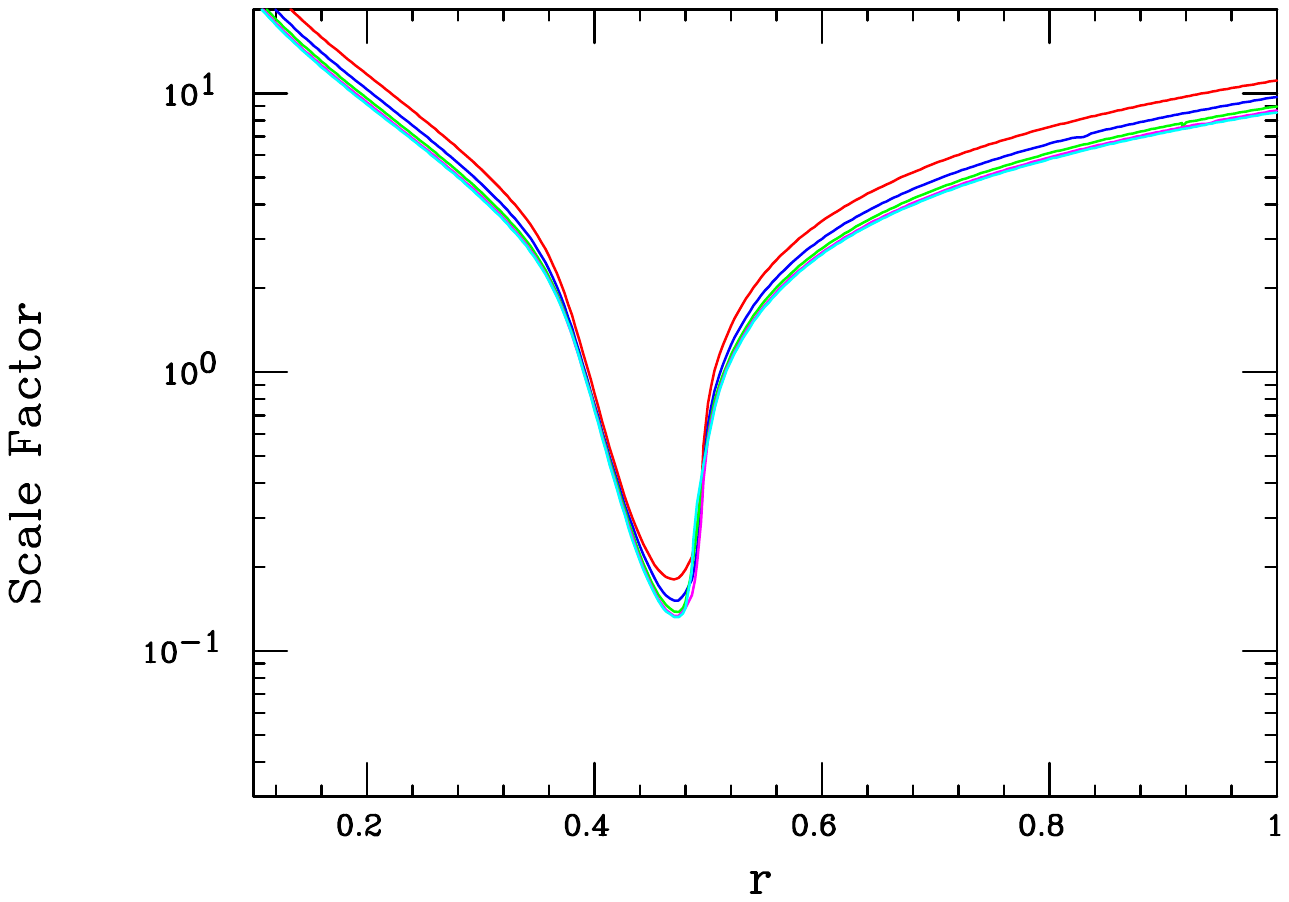}
\hspace*{-1.9cm}
\includegraphics[width=4.2in,angle=0]{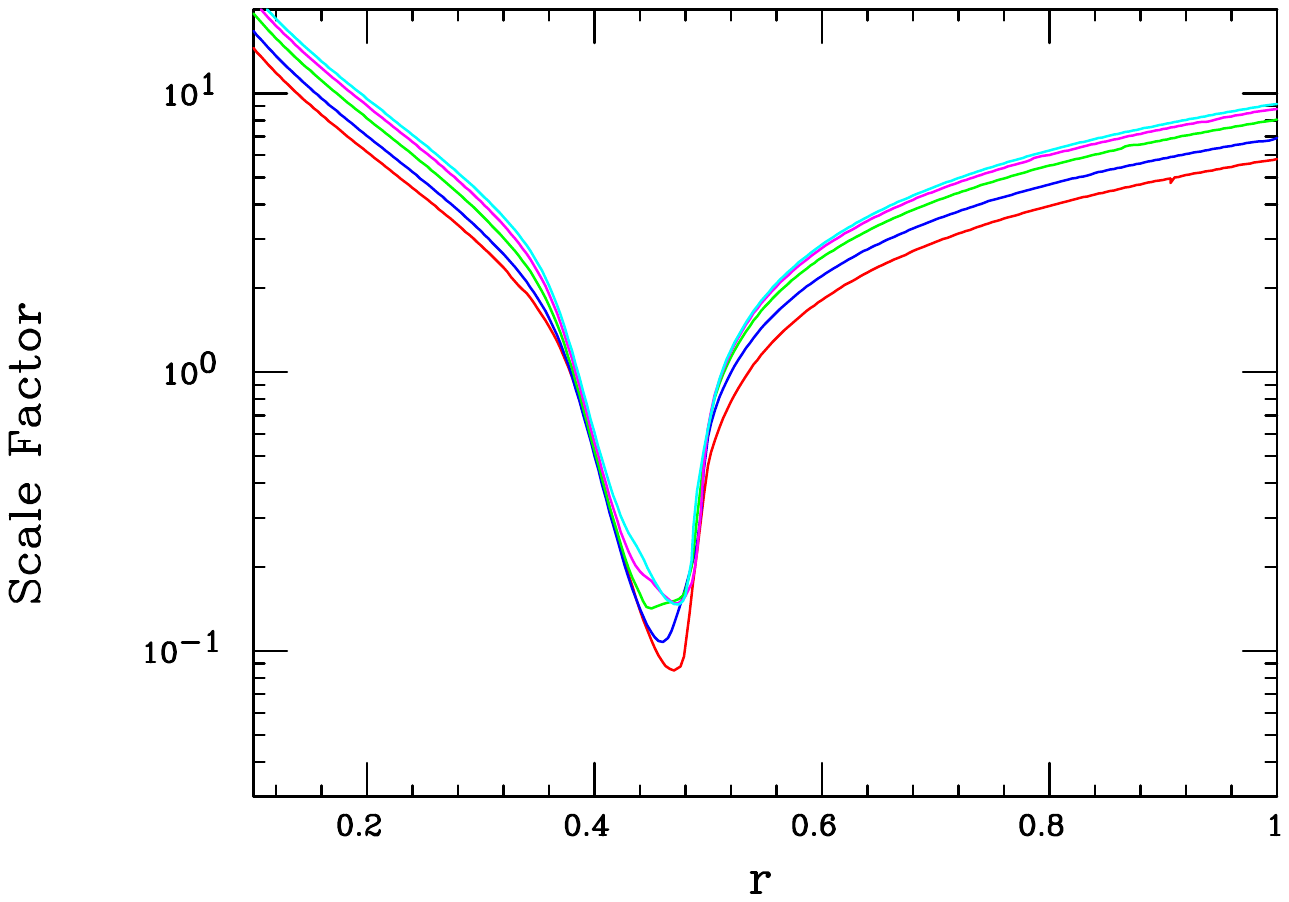}}
\vspace*{-1.00cm}
\centerline{\includegraphics[width=4.2in,angle=0]{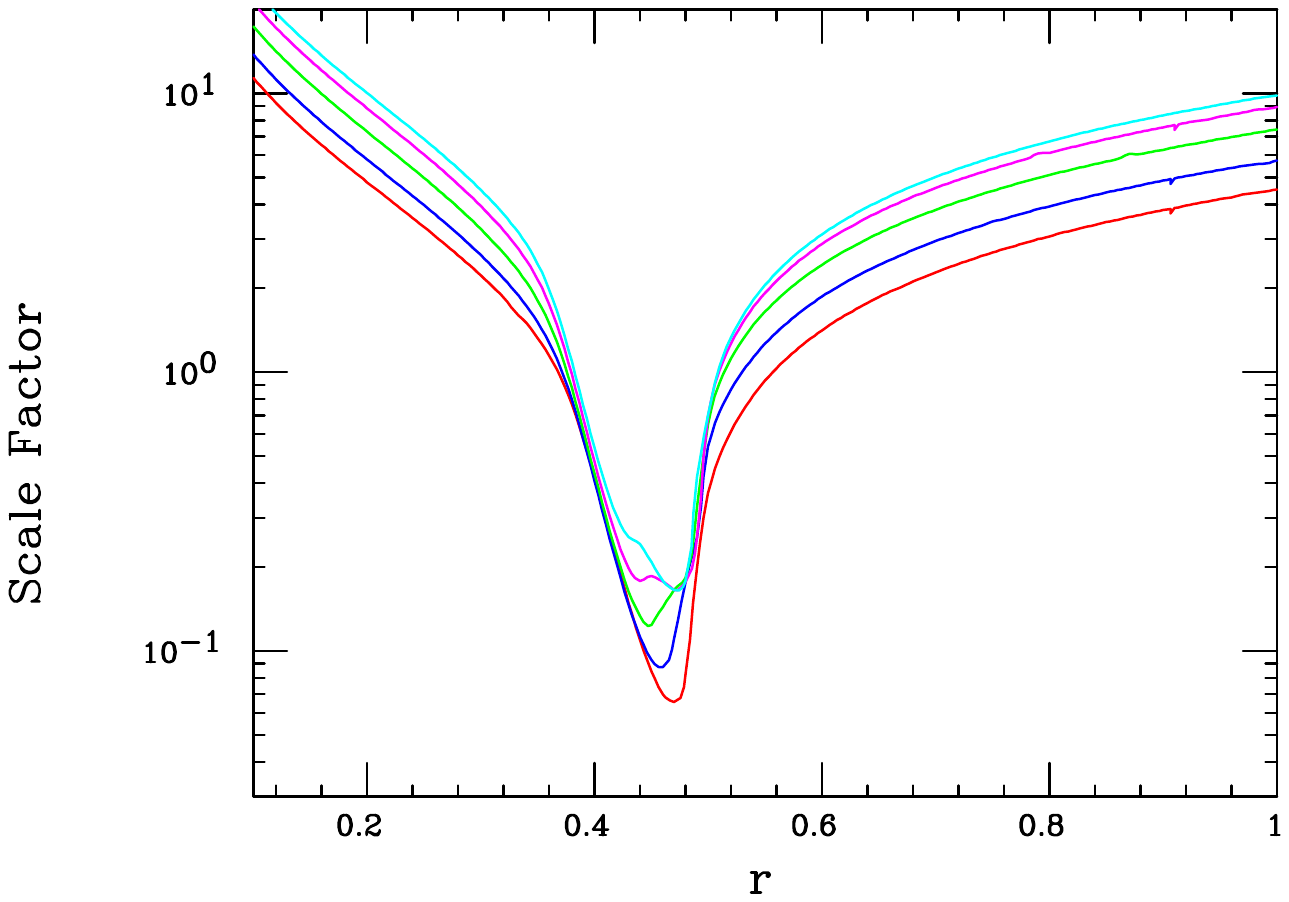}
\hspace*{-1.9cm}
\includegraphics[width=4.2in,angle=0]{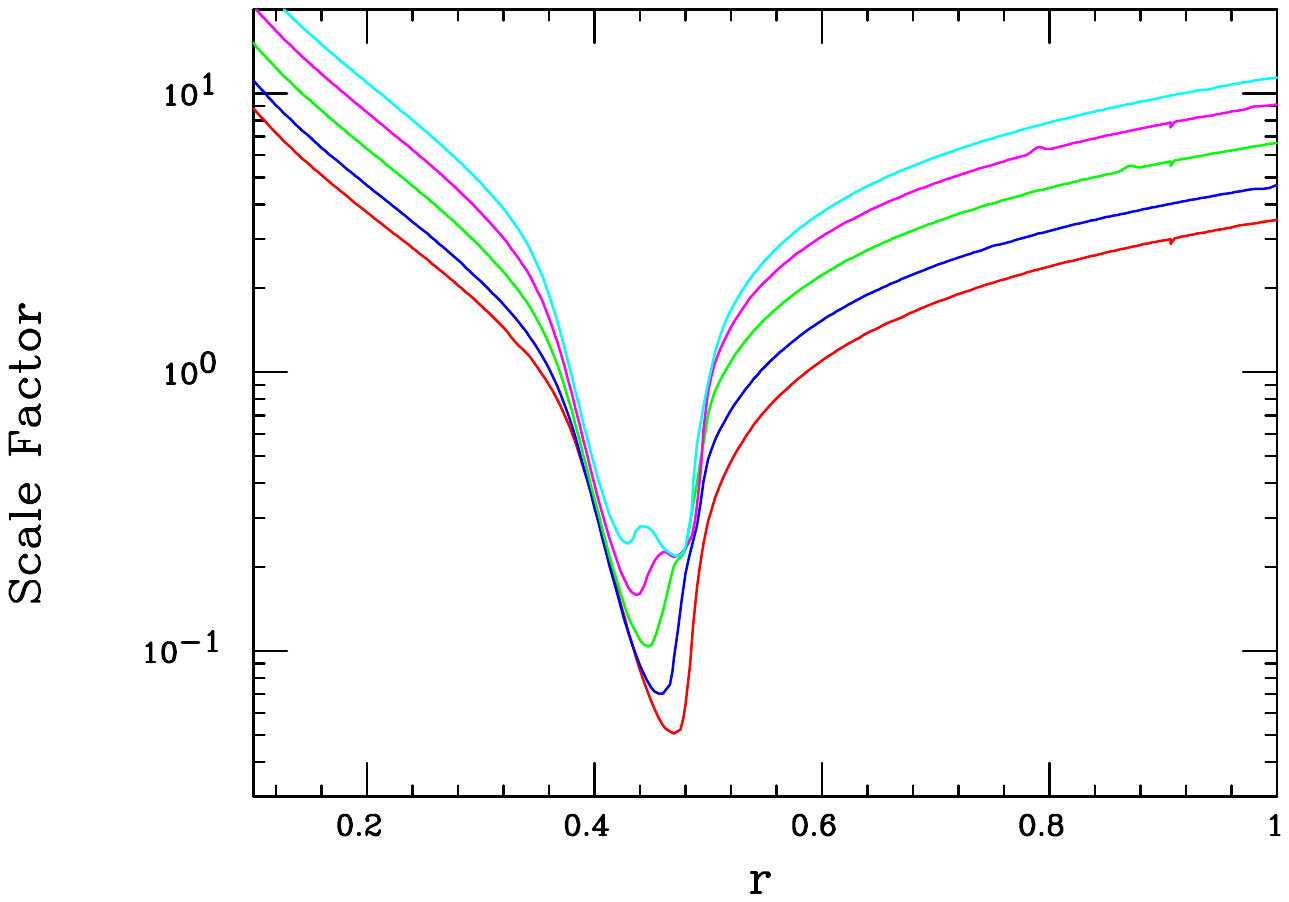}}
\vspace*{-0.60cm}
\caption{The numerical value of the scale factor ${\cal F}$ required to obtain the observed relic density as a function of $r$ assuming that $\delta=0.05$ (red), 0.10 (blue), 0.15 (green), 0.20 (magenta) 
or 0.25 (cyan).  In the (Top Left) panel, $x=0$, (Top Right) $x=0.05$, (Bottom Left) $x=0.10$, (Bottom Right) $x=0.20$, respectively. Also as noted in the text, the recent null searches for pairs of 
lepton-jets arising from decays of $H(125)$\cite{ATLAS:2024zxk,CMS:2024jyb} strongly suggest that $r<1/2$ {\it unless} the mixing of the SM Higgs with $h_D$ is very highly constrained. }
\label{fig1}
\end{figure}

We now briefly consider the direct detection of DM in this setup. The effect of mixing here is relatively minor if the higher mass state, $\chi_2$, cannot be excited at tree-level. 
In the mass region of interest, it is mostly the scattering of DM off of (bound) electrons that provides a constraint on this model class; here, we see that both elastic $\chi_1 e \to \chi_1 e$ and inelastic 
$\chi_1 e \to \chi_2 e$ processes are, in principle, of relevance depending upon the value of $x=\sin ^2\theta$. However, for the range of $\delta \gsim 0.05$ that we consider, unless the DM 
is somehow highly boosted (see, \eg, \cite{Ilie:2010vg,Agashe:2014yua,Giudice:2017zke}), the inelastic process is kinematically forbidden as the DM has insufficient energy to excite the more 
massive $\chi_2$ state. Thus when $\sin^2 2\theta$ is close to unity, the only kinematically allowed process will remain elastic scattering and this will only be allowed to occur at the 1-loop level (via 
a virtual $\chi_2$) and so be quite suppressed\cite{Bell:2018zra} by many of magnitude, becoming essentially invisible.

\begin{figure}[htbp]
\centerline{\includegraphics[width=4.2in,angle=0]{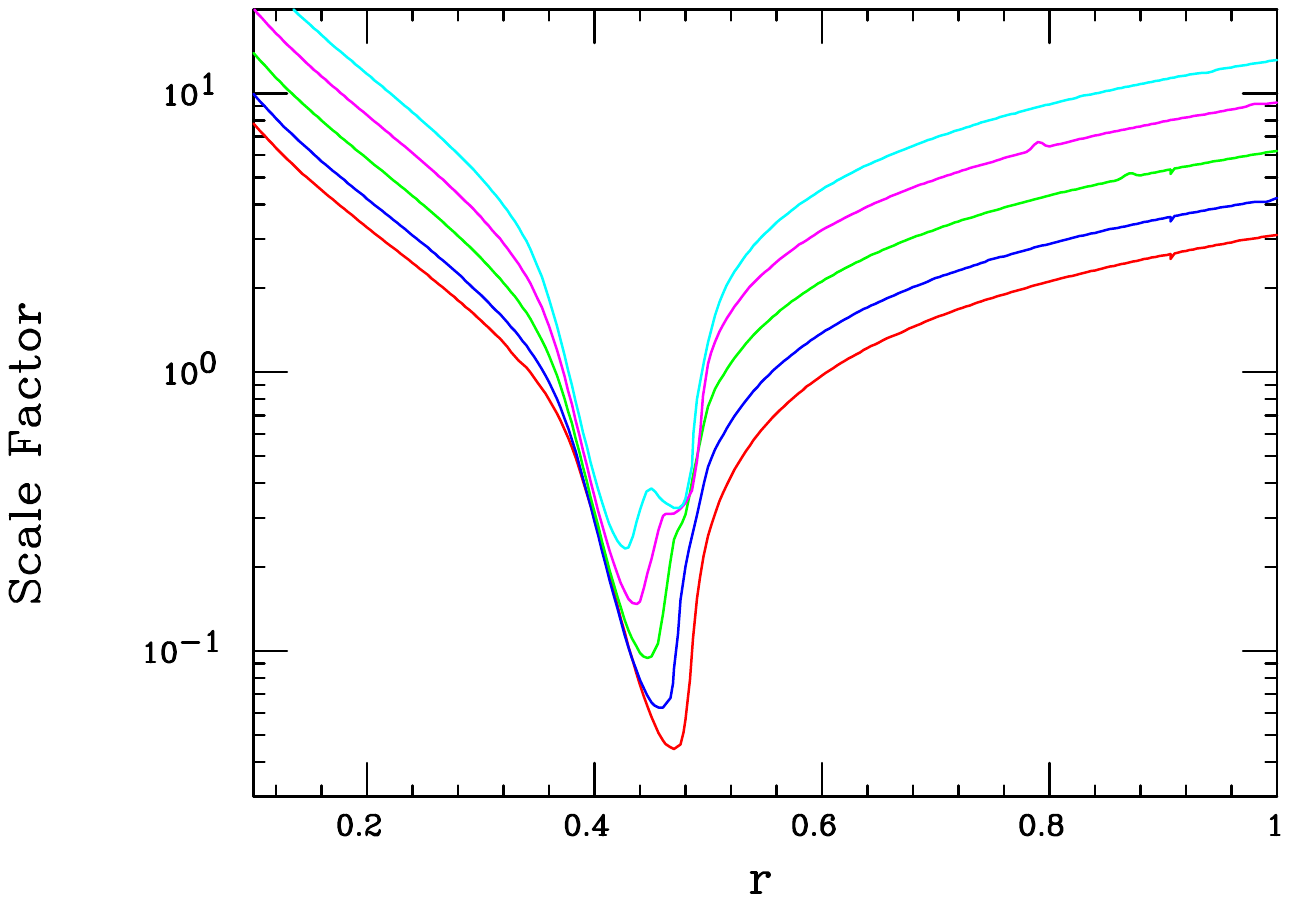}
\hspace*{-1.9cm}
\includegraphics[width=4.2in,angle=0]{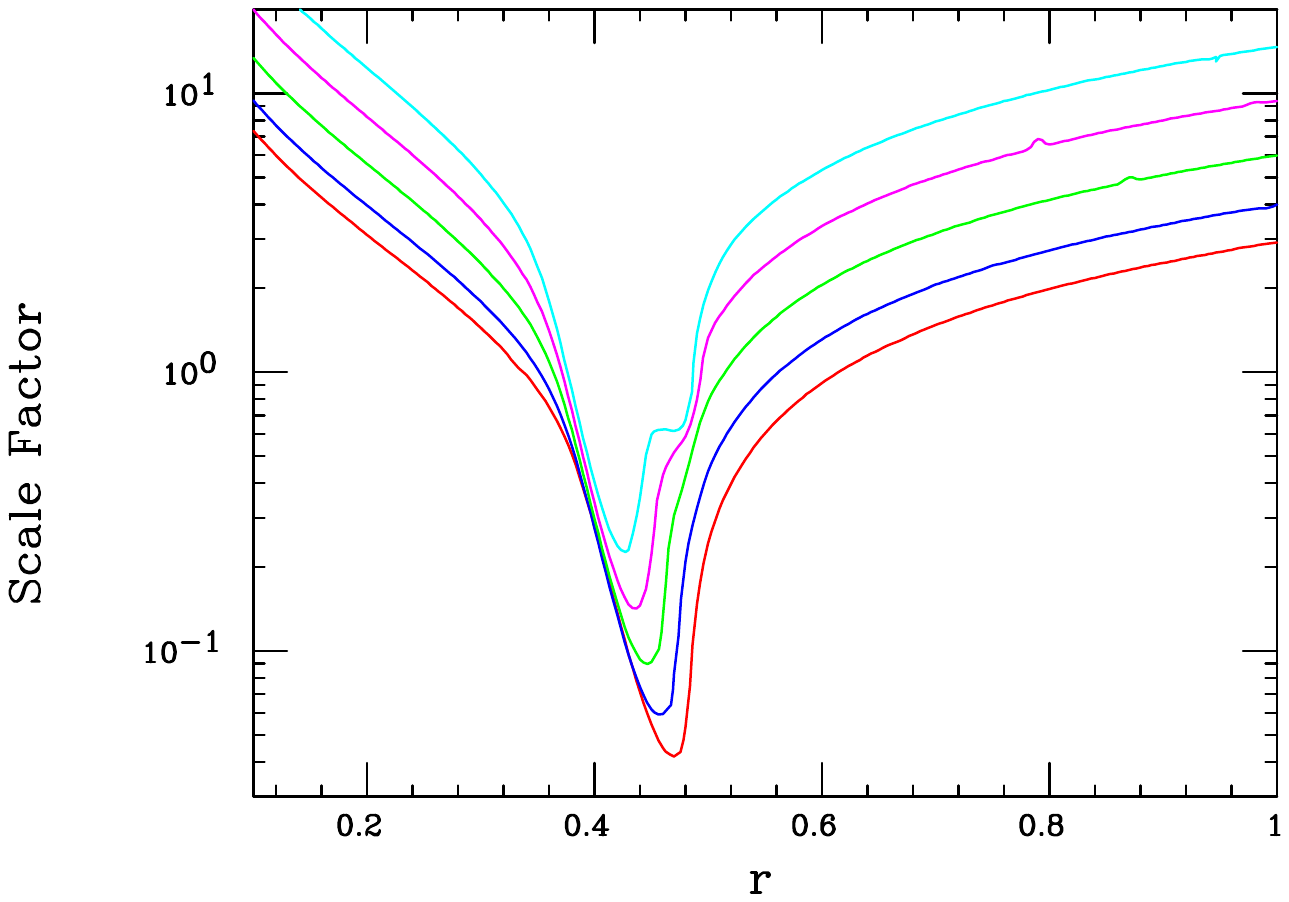}}
\vspace*{-1.00cm}
\centerline{\includegraphics[width=4.2in,angle=0]{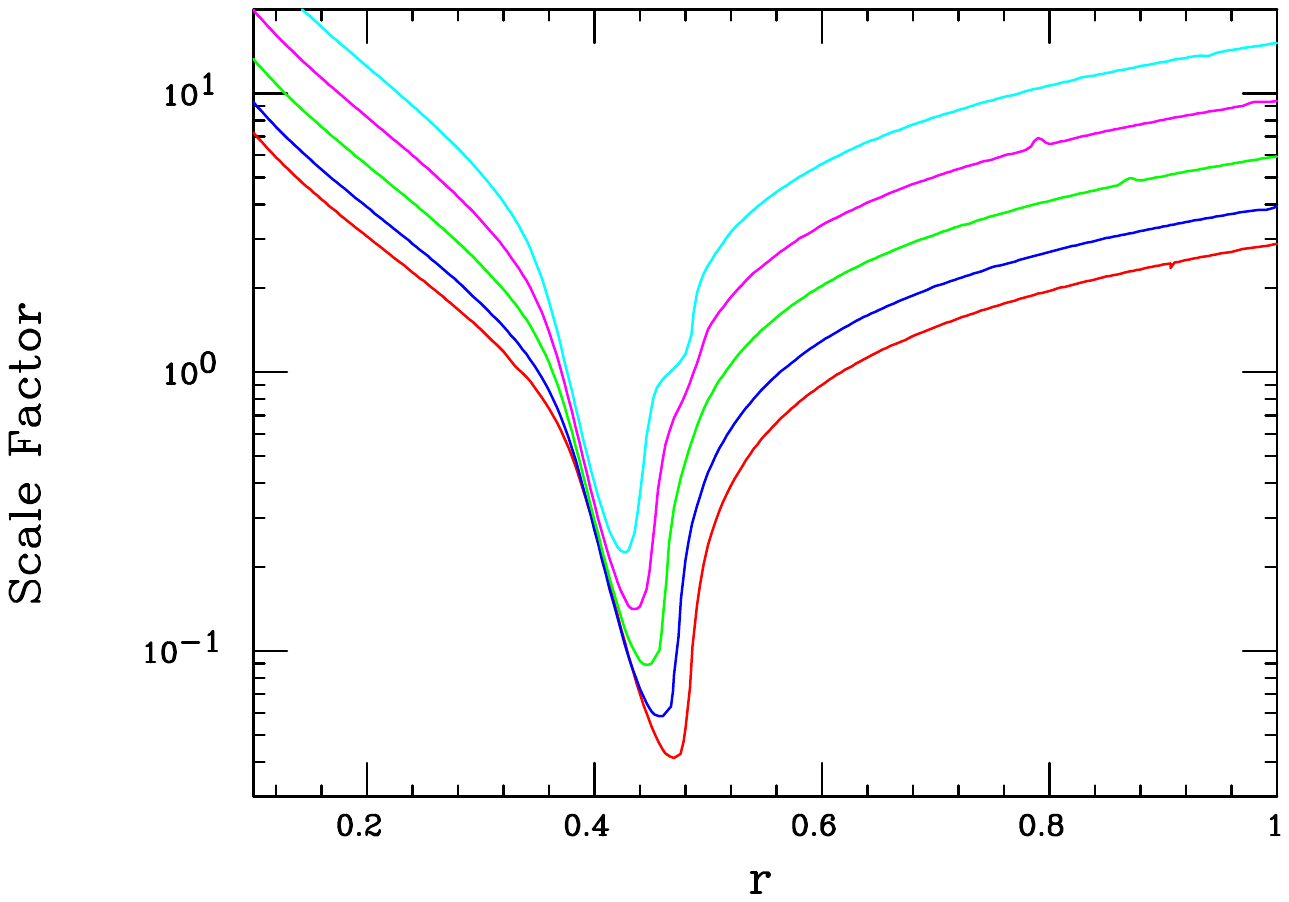}
\hspace*{-1.9cm}
\includegraphics[width=4.2in,angle=0]{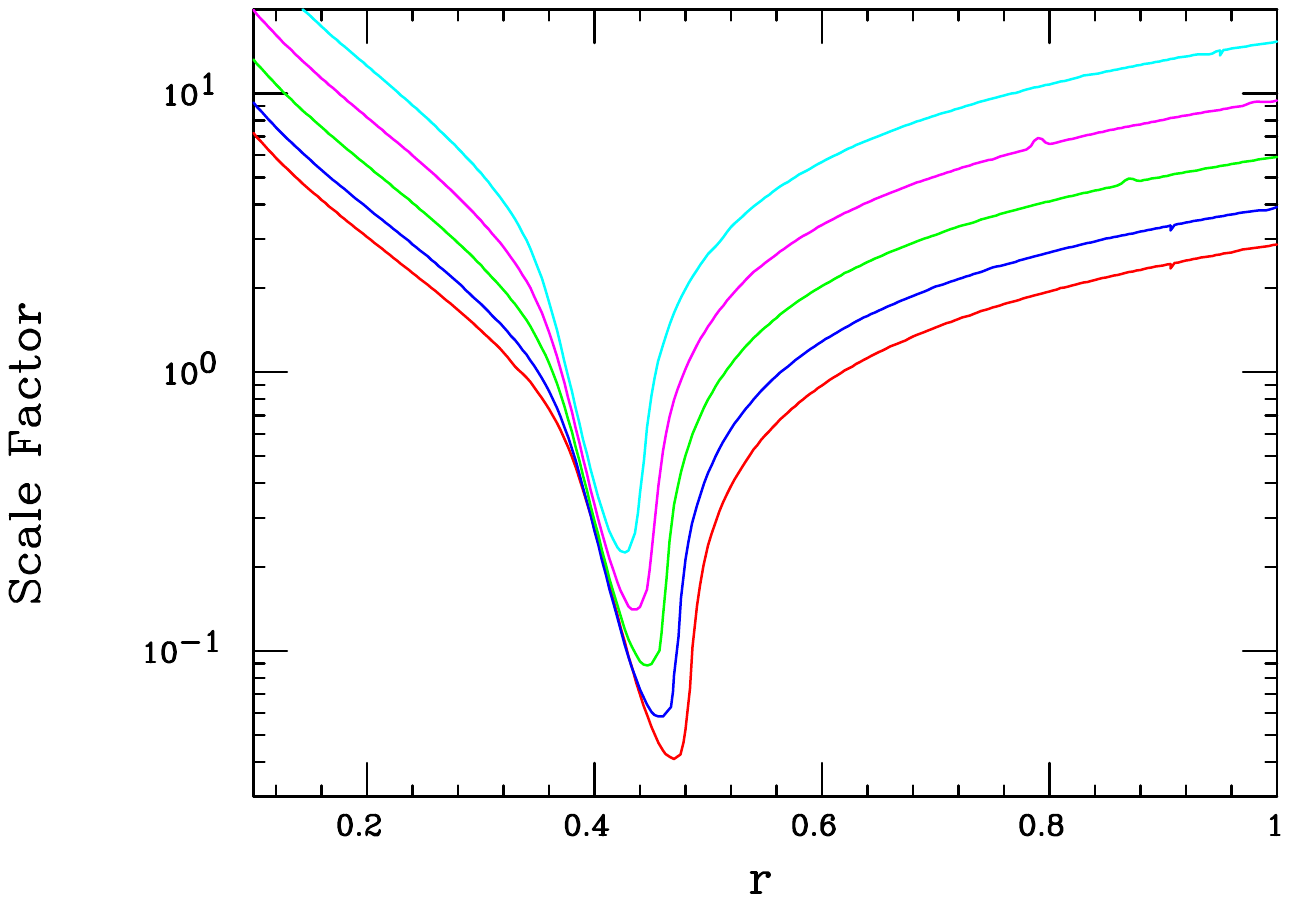}}
\vspace*{-0.60cm}
\caption{Same as in the previous Figure but now in the (Top Left) panel, $x=0.30$, (Top Right) $x=0.40$, (Bottom Left) $x=0.45$, (Bottom Right) $x=0.50$, respectively. }
\label{fig2}
\end{figure}

In this setup, the DM scattering cross section is then little changed from the usual result apart from the additional weighting due to mixing angle effects. For a {\it free} electron at tree-level, the 
$\chi_1-e$ cross section is then given by
\begin{equation} 
\sigma_e = \frac{4\alpha \mu^2 g_D^2 \epsilon^2}{m_V^4} ~(1-2x)^2\,
\end{equation}
where $\mu=m_em_\chi/(m_e+m_\chi)$ is the reduced mass $\simeq m_e$ for the DM masses $\sim $100 MeV of interest to us here. Numerically, one finds, in terms of ${\cal F}$, that  
\begin{equation} 
\sigma_e \simeq  3.0\cdot 10^{-40}~{\rm{cm}^2} ~\Big(\frac{100 ~{\rm {MeV}}}{m_V}\Big)^2 ~{\cal F}^2  ~ (1-2x)^2\,
\end{equation}
values which are not too far away from the bounds obtained by current experiments\cite{PandaX-II:2021nsg,DarkSide:2022knj,XENON:2019gfn} and which vanishes completely at tree-level 
in the pure pseudo-Dirac case. Clearly, smaller values of ${\cal F}$ and/or larger values of $m_V$ would provide us with more freedom from these currently existing constraints.

\section{The Dark Higgs Sector}

Going beyond the previously considered complex singlet DM setup and recalling the fact that light Dirac DM in the KM scenario is either excluded or at best highly disfavored, the doublet DM 
scenario discussed 
above was seen to require a somewhat more complex symmetry breaking sector. Given the toy model discussion within the $G_D=SU(2)_I\times U(1)_{Y_I}$ context here, the most minimal dark 
sector Higgs fields (in addition to that of the usual $G_D$ singlet, SM Higgs doublet, $\Phi$) which are necessarily all SM singlets, are then just the doublet, $H_D$, with $Y_I=1$, whose large 
$\gsim 10$ TeV vev, $v_D$, is responsible for high scale $G_D\to U(1)_D$ breaking as well as for the PM masses, and $T$, the $Y_I=2$ triplet that generates both the DM Majorana mass splittings 
as well as the DP mass due to the $U(1)_D$ breaking at a scale $v_T\sim 0.1-1$ GeV.  In such a setup the full Higgs potential is just the sum
\begin{equation} 
V=V_{Dark}+V_{SM}\,
\end{equation}
with the potential for the Higgs fields in this dark sector being described by (here employing the familiar $2\times 2$ matrix notation for the triplet, $T$), 
\begin{eqnarray}
V_{Dark}=m^2H_D^\dagger H_D+\lambda_1(H_D^\dagger H_D)^2+M^2 Tr(T^\dagger T)+\lambda_2 [Tr(T^\dagger T)]^2+\lambda_3 Tr[(T^\dagger T)^2]\nonumber\\
~+\lambda_4 (H_D^\dagger H_D) Tr(T^\dagger T)+\lambda_5 H_D^\dagger (TT^\dagger) H_D+\Big(\mu H_D^T~ i\tau_2 T^\dagger H_D+{\rm h.c.}\Big)\,,
 \end{eqnarray} 
where $\tau_2$ is just the familiar Pauli spin matrix and we can express the individual field components of $H_D$ and $T$, all in $SU(2)_I$-space, as (here the various indices reflect the values 
of $Q_D$)
\begin{equation}
T=\begin{pmatrix} \frac{T_1}{\sqrt 2} & T_2=\frac{v_T+h_D+i\eta}{\sqrt 2} \\ T_0 & -\frac{T_1}{\sqrt 2} \\  \end{pmatrix},~~~~~ H_D=\begin{pmatrix} \tilde h \\ h_0=\frac{v_D+h_D'+ia}{\sqrt 2}\\ \end{pmatrix} \,,
\end{equation}
and where $v_T \simeq 0.1-1$ GeV and $v_D\gsim 10$ TeV are the two vevs that have already been encountered above. The remaining part of the potential involving the SM Higgs, $\Phi$, is then 
\begin{equation}
V_{SM}=m_{SM}^2 \Phi^\dagger \Phi+\lambda_{SM}(\Phi^\dagger \Phi)^2+\lambda' (\Phi^\dagger \Phi)(H_D^\dagger H_D) +\lambda'' (\Phi^\dagger \Phi)Tr(T^\dagger T)\,.
 \end{equation} 

Since $v_T/v_D \sim 10^{-(4-5)}$, $v_{SM}/v_D \sim 10^{-2}$ and $v_T/v_{SM} \sim 10^{-(2-3)}$, in what follows we can (most of the time) work quite safely to leading order in (generally the 
squares of) these 
small vev ratios{\footnote{In what follows, we will also neglect the possible effects of CP-violation for simplicity.}}.  Then the complex field $\tilde h$, which has $Q_D=1$, becomes the eaten 
Goldstone boson for the $W_I$, while $\eta$ and $a$, which both have $Q_D=0$, together become the eaten Goldstones for the $Z_I$ and $V$, still leaving us with six degrees of freedom for 
the physical dark Higgs fields remaining. Similarly, in this limit, three of the fields in $\Phi$ become the Goldstone modes for the $W$ and $Z$ as usual. Thus one sees, as expected, that $h_D'$ 
will play essentially the same role for $G_D$ as does $h_{SM}$ within the familiar SM framework. Now as we observed during the relic density discussion above, the limit on the invisible branching 
fraction of the $H(125)$, $B_{inv} \lsim 0.1$, very strongly constrains any mixing between the SM and dark sector Higgs fields which can dominantly couple to purely invisible modes that are 
kinematically accessible in $H(125)$ decay. As we saw, since the coupling of $h_D$ to $2V$ is large this mixing must be highly suppressed. Here, $h_D$, is a member of the triplet, $T$, 
so this constraint applies to $\lambda''$ quartic; in particular $h_{SM}-h_D$ mixing in this case found to be roughly proportional to the ratio $\simeq \lambda'' v_Tv_{SM}/m_{h_{SM}}^2$ and 
so $\lambda''$ must be rather small.  

Now similarly, the corresponding magnitude of the $h_D'-h_{SM}$ mixing is found to be roughly given by a similar ratio, $\sim \lambda' v_{SM}v_D/m_{h_D'}^2$, and so one can imagine 
at least {\it approximately} setting both $\lambda', \lambda'' \to 0$ and ignoring (or the moment) any coupling of the SM Higgs with those in the dark Higgs sector in what follows. 
However, while it is easily seen that the restriction on $\lambda''$ is clearly quite strong due to the invisible branching fraction constraint, a similar analogous argument cannot obviously be made for 
$\lambda'$ as $h_D'$ (in the guise of the $h_{1,2}$ mass eigenstates below) will couple predominantly only to the heavy states, \eg, $W_IW_I^\dagger$, $2Z_I$ and pairs of PM fermions 
to which it gives mass, and not to, \eg, $2V$. None of these potential final states are kinematically accessible in the decay of $H(125)$. In such a case the corresponding small mixing with 
$h_{SM}$ will then simply lead to a suppression of all of the $H(125)$ SM partial widths by a common overall factor{\footnote {the so-called '$\cos \theta$' effect}} of order a 
few percent at most, but will leave the branching fractions invariant. If such a mixing were to be at all significant, a further possible bound on its magnitude then arises from (delayed) unitarity 
requirements \cite{Kang:2013zba,Nagai:2014cua} via the standard arguments and which suggests that the corresponding mixing angle is roughly bounded from above by 
$\simeq 0.2 \cdot (3.5~ {\rm TeV})/{\rm min}(m_{h_{1,2}})$.  More generally, it is to be noted that finite mixing effects of such a magnitude, even when they do occur, will do little to shift the SM 
Higgs mass (in terms of, \eg, the vevs $v_{T,D}$) as such mass shifts are found to be higher order in the ratios of these vevs. Only the possible decay modes of H(125), as we've been 
discussing, will be modified to leading order. 

Returning to the scalar potential above, as usual, the mass-squared parameters $m^2,M^2$ and $m^2_{SM}$ appearing there can be eliminated via the minimization conditions. Then the  
resulting physical fields will obtain the following approximate squared masses to leading order in the small vev ratios:
\begin{equation} 
m_{h_{SM}}^2 \simeq 2\lambda_{SM}v_{SM}^2, ~~m_{T_1}^2 \simeq \frac{1}{4}\lambda_5 v_D^2, ~~m_{I_0}^2\simeq \frac{1}{2}\lambda_5v_D^2,~~m_{h_D}^2\simeq 2(\lambda_2+\lambda_3)v_T^2
\,,
\end{equation}
where $T_1$ remains a complex field, $T_1^*= T_{-1}$, and we have decomposed the complex, $Q_D=0$ field $T_0$ as $T_0=(R_0+iI_0)/\sqrt 2$ with $R_0(I_0)$ being CP-even (odd). Then 
we see that the remaining two CP-even fields with $Q_D=0$, $R_0$ and $h_D'$, will mix when $\mu \neq 0$ forming a $2\times 2$ mass-squared matrix in the $h_D'-R_0$ basis which in the 
same limit of the small vev ratios discussed above is given by: 
\begin{equation}
{\cal M}^2_{h_D'R_0}\simeq  \begin{pmatrix} 2\lambda_1 v_D^2 & \frac{-\mu v_D}{\sqrt 2} \\  \frac{-\mu v_D}{\sqrt 2}& \frac{1}{2}\lambda_5 v_D^2\\  \end{pmatrix}\,,
\end{equation}
which is diagonalized via a simple rotation with a mixing angle given by 
\begin{equation}
\tan 2\kappa = \frac{-\sqrt 2 \mu}{(2\lambda_1-\lambda_5/2)v_D}\,.
\end{equation}
Clearly if $\mu \sim v_D$ this mixing angle, $\kappa$, can be seen to be of order unity assuming that $\lambda_1$ and $\lambda_5$ are of comparable magnitude. We will refer to these two mass 
eigenstates as $h_{1,2}$, \ie, $R_0=h_1c_\kappa-h_2s_\kappa$, \etc, having the masses $m_{h_{1,2}}$, with $m_{h_1}<m_{h_2}$, respectively, in the analysis below.

From this discussion, we see that the present setup is manifestly quite different from the case where the DM is a complex $G_D$ singlet. There, after SSB, only the two CP-even dark Higgs fields, the 
analogs of $h_D,h_D'$, will remain in the physical spectrum in addition to the usual SM Higgs. Here we see a necessarily more complex situation: in addition to those fields, assuming that the DM is 
Majorana/pseudo-Dirac, we require also the existence of the complex field, $T_1$, the CP-odd field, $I_0$, as well as the extra CP-even field, $R_0$, which in general undergoes an $O(1)$ mixing 
with $h_D'$. Since these additional scalar fields all obtain masses via the large $G_D$-breaking vev, $v_D$, they (together with the DM's heavy fermionic partner, $\psi$) may provide unique 
signatures for the present setup at the LHC or at future colliders to which we now turn.

\section{Collider Phenomenology: The Production and Decays of New States}

Without knowing the values of all of parameters in this setup, we can at best only make a few general, semi-quantitative comments on the mechanisms available for dark Higgs production. 
If one completely neglects the small vev ratios above, then the dark Higgs sector, as far as the scalar potential-induced interactions are concerned, becomes totally isolated from the SM and 
the various physical states will only interact with each other as well as with the $G_D$ gauge bosons{\footnote {In the discussion that follows, we will also neglect any very small mixing, 
$\sim 10^{-4}$, between the SM $Z$  and the $Z_I$.}}. Of course, the heavy $W_I,Z_I$ gauge fields themselves {\it do} interact with (some of) the SM fields even in the limit when $\epsilon \to 0$ 
as these SM fields will share common representations of $G_D$ with PM, \eg, $(\nu_L,e_L)$ with $(N_L,E_L)$ and $d_R$ with $D_R$ in the simplest $E_6$-inspired toy model discussed above 
and in earlier work. In this approximate limit where the mixing between the SM Higgs and the dark Higgs fields are completely ignored, these dark states will also be pure SM singlets which leads 
to a number of immediate consequences. In such a case, there are only a few ways that these dark sector Higgs fields may be produced at a collider with respectable rates. Of course, DP 
exchange reactions are aways present between 
the charged SM fields and any fields with $Q_D\neq 0$, but these can generally (except is special circumstances) be ignored as they lead to $\epsilon^2$ suppressed cross sections. However, 
since the heavy $G_D$ gauge fields, $W_I,Z_I$, will couple to (at least some of) the SM fermions, generically denoted as $f$, without any associated suppression factors (\eg, $\epsilon$), they can 
be employed to produce these new dark Higgs in analogy with the charged and heavy neutral Higgs bosons in THDM setups. An obvious possibility is the pair-production of these new heavy states 
via their $G_D$ gauge couplings, \eg, via $Z_I$ exchange, \eg, $\bar f f \to Z_I^{(*)} \to I_0 h_{1,2}, T_1T_{-1}$, where, depending upon the various relative masses, the $Z_I$ may or may not be 
on-shell and resonant. For example, if both $\lambda_5 \simeq \lambda_{SM}$ and $g_I/c_I \simeq g/c_w$, then $Z_I \to T_1T_{-1}$ would be an allowed on-shell decay mode{\footnote {Similarly, 
the decay $W_I\to T_1h_D$ would also be kinematically allowed.}}. This would be quite advantageous as this rate would be significantly resonantly enhanced and without which, as we'll discuss 
below, rather tiny cross sections for this type of $Z_I$-mediated process would be obtained.

Another possibility to access these new states is associated production, analogous to of $Zh_{SM}$ in the SM; however, such processes are in most cases generally expected to be sub-leading 
in the present setup since the relevant couplings are generated via their associated vevs and here $v_T$ is quite small. Thus while the $\bar f f\to Z_Ih_D'$ production amplitude is proportional to 
$v_D$ and so might yield a significant cross section (that will be discussed below), the analogous amplitude for $\bar f f\to Z_IR_0$ is instead proportional to $v_T$ and thus is relatively 
suppressed by a very large factor. Fortunately, both $h_{1,2}$ may have $h_D'$ components although the lighter state $h_1$'s coupling is expected to be somewhat suppressed.  Thus both of 
these modes might be accessible except for the additional phase space suppression in the case of $h_2$. However, as this is, by definition, a non-resonant process one would need to pay the large 
price of having to produce (through electroweak strength interactions) a pair of very heavy particles, each with a mass $\gsim$ a few TeV or more, which will only be possible at FCC-hh as we will 
later below. In fact, one finds that the relevant cross sections are generally small.

Now in the THDM case, we recall that $W^\pm$ exchange plays a rather important role in providing possible access to such final states as $H^\pm(H,A)$; however, the $W_I$ is unfortunately not 
able to 
play the analogous role for us here. Recall that since the $W_I$ couples a SM fermion, $f$, to a corresponding PM field with the same electroweak and strong properties, $F$, it cannot be singly 
produced{\footnote {Note that since $W_I^{(\dagger)}$ carries $|Q_D|=1$, it can only be pair-produced or created in association with a PM field assuming a purely SM initial state.}} or exchanged 
in the $s$-channel in the limit that $v_T,\epsilon \to 0$ and so is essentially irrelevant for the production of these new heavy dark Higgs states. 

In order to make more firm predictions than the semi-quantitative statements above, it is clear that we need to substantially collapse the large parameter space of this model to make it more tractable.  
For example, as one can tell from the previous discussion(s), the wealth of model parameters makes a fully detailed understanding of the interplay between the new heavy scalar and corresponding 
gauge sectors 
in all generality essentially impossible. To this end, and for demonstration purposes, let us consider a not all too completely implausible toy benchmark scenario partially alluded to above and having 
been employed for numerical purposes in some earlier work:  ($i$) We imagine that all of quartic couplings appearing in $V_{Dark}$ take on the same common value and are equal to 
$\lambda_{SM}$ and also that $\mu=\lambda_{SM}v_D$. ($ii$) We also assume that both of the $SU(2)_I$ and $U(1)_{Y_I}$ gauge couplings, $g_I,g_I'$,  are identical to their corresponding SM 
counterparts, $g,g'$. In such a setup, clearly all on the various interactions within the purely dark 
sector are now known - but we actually know more than this as the $h_D'-R_0$ mixing angle also become determined ($s_\kappa=-0.369,c_\kappa=0.929$) and the masses of all these new heavy 
states are also fixed up to an {\it overall} scale, \ie, $v_D$. Under these assumptions, Table 1 presents the scaled masses of the heavy dark Higgs fields and those of the new $G_D$ gauge bosons 
in units of $v_D$. Note that the masses of PM fields as well as $\psi$ are unfortunately {\it not} fixed by these assumptions. We may still ask, however, what is the lower bound on the value of 
$v_D$ allowed by current experiment? 

\begin{table}
\caption{Dark Gauge and Higgs Sector Masses}\label{higgstab}
The approximate masses of the heavy gauge and dark sector scalar fields in units of the dark Higgs doublet vev, $v_D$, for the toy benchmark model point discussed in the text.
\begin{center}
\begin{tabular}{ l c }
\hline
State & Particle Mass in Units of $v_D$   \vspace{0.1cm}\\
\hline
$W_I$ &0.326 \\
$Z_I$ &  0.370\\
$T_1$ & 0.177 \\
$I_0$ & 0.254 \\
$h_1$ & 0.168 \\ 
$h_2$ & 0.543 \\
\hline 
\end{tabular}
\end{center}
\end{table} 

Since the $Z_I$ plays an important role in the production of the new scalar states, we can directly use these benchmark assumptions to learn something about its mass which subsequently will 
constrain the value of $v_D$ from below. Assuming that $Z_I$ can only decay on-shell to SM final states (a fairly good first approximation as we'll see), we can determine its dilepton signal 
cross section at the 13 TeV LHC employing the narrow width approximation (NWA), \ie, $\sigma B(Z_I\to l^+l^-)$, and compare it to the experimental search limits\cite{ATLAS:2019erb,CMS:2021ctt}.  
Employing the 139 fb$^{-1}$ ATLAS results\cite{ATLAS:2019erb}, one finds that $M_{Z_I}>5.0$ TeV, not far from that obtained for a heavy sequential $Z'$ with SM couplings, $Z_{SSM}$, of 5.1 
TeV; these production rates at the 13 TeV LHC are shown in the top panel of Fig.~\ref{fig3}. Given the relationship between $Z_I$ and $v_D$, this bound tells us that $v_D > 13.5$ TeV and thus, 
with our benchmark assumptions, provides a lower limit on the $W_I$ mass as well as the masses of all the new dark scalar states. For completeness we note that a corresponding 
null search at the 14 TeV LHC with 3 ab$^{-1}$ of integrated luminosity would imply that $M_{Z_I}>5.8$ TeV. On the other hand, the 100 TeV FCC-hh with an integrated luminosity of 30 ab$^{-1}$ 
would allow us to probe for a $Z_I$, under these same assumptions,  with a mass as high as $\simeq 37.5$ TeV as is shown in the lower panel of Fig.~\ref{fig3} by employing the analysis in 
Ref.\cite{Helsens:2019bfw}. However, a $Z_I$ with a mass close to this search reach would not likely yield a usable number of dark Higgs fields to allow for any further study at either the LHC or 
FCC-hh. It is important to note before continuing that if the $Z_I$ {\it does} decay to non-SM final states, thus reducing the leptonic branching fraction by, say, a factor of 2, the resulting bound on 
the $Z_I$ mass, and hence on $v_D$, would only be slightly reduced by $\sim$ a few percent due to the behavior of the fall-off of the cross section with mass as can be seen in the Figure. 

\begin{figure}[htbp]
\centerline{\includegraphics[width=5.0in,angle=0]{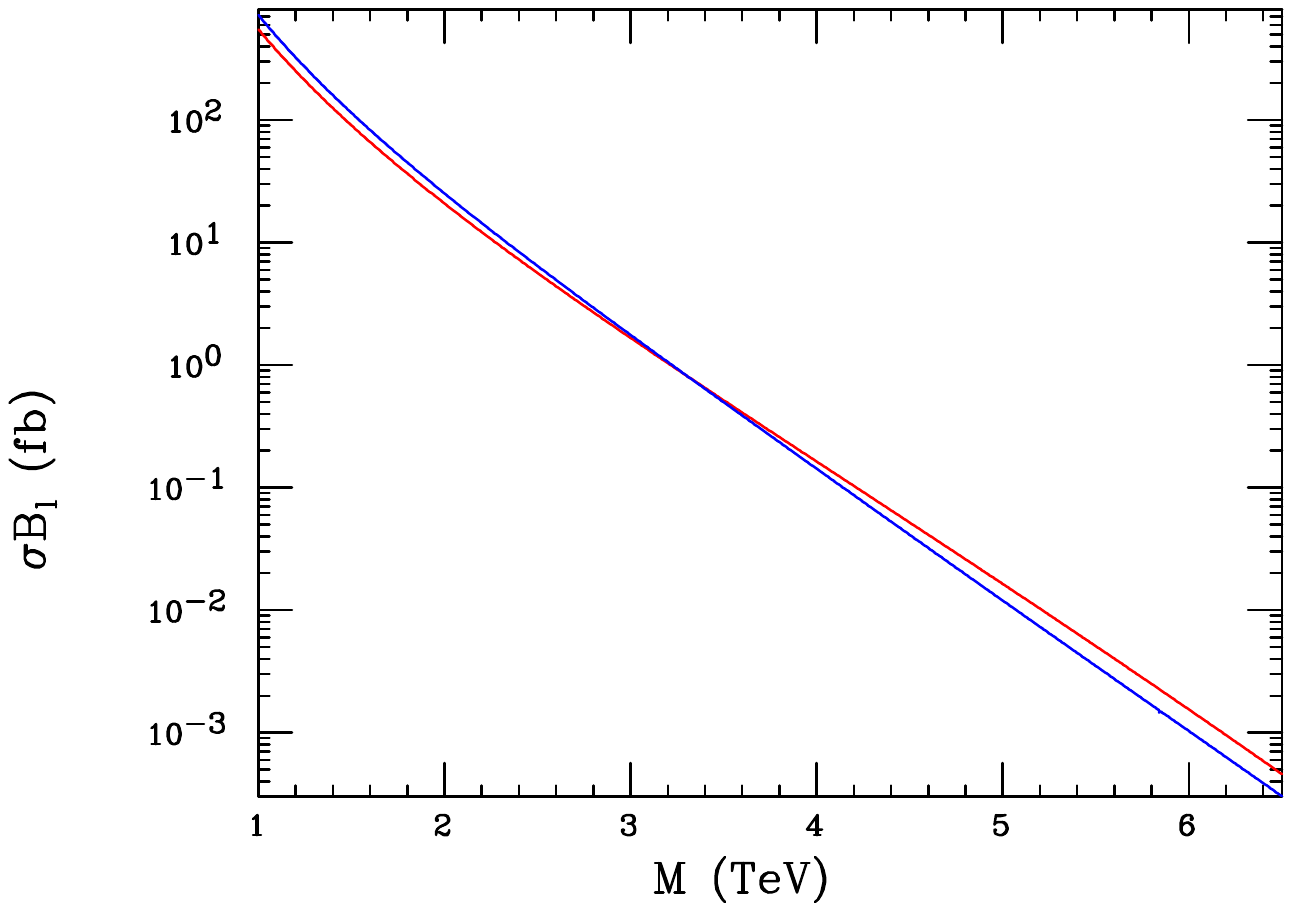}}
\vspace*{-1.8cm}
\centerline{\includegraphics[width=5.0in,angle=0]{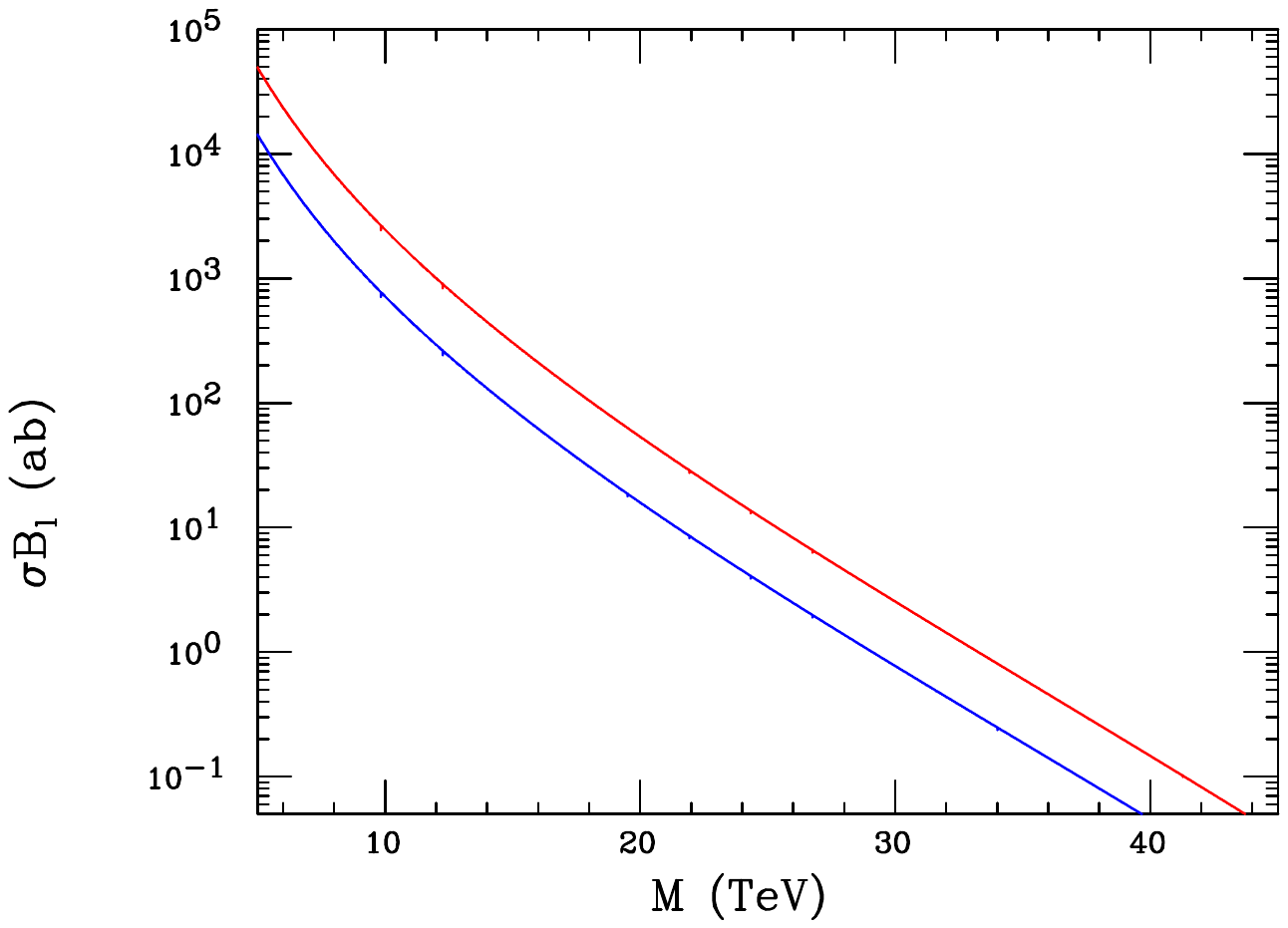}}
\vspace*{-1.3cm}
\caption{Comparison of the cross section times leptonic branching fractions for the production of the new neutral gauge bosons $Z_{SSM}$ (red) and $Z_I$ (blue) in the narrow width 
approximation,  as discussed in the text, (Top) for the 13 TeV LHC and (Bottom) for the 100 TeV FCC-hh. }
\label{fig3}
\end{figure}

Since $\lambda_{SM} \simeq 0.129$ is relatively small, one might imagine that such a judicious choice would make the new scalar states light and so easier to discover; looking at the entries in 
Table 1, we see that unfortunately this may not be the case. The just discussed production of $Z_I$ {\it can} give us access to the dark Higgs states that can be produced on-shell in $Z_I$ 
decay taking advantage of the resonant enhancement while still paying the price for the production of a pair of heavy states.  Unfortunately, with our benchmark model assumptions, the only 
on-shell decay of the $Z_I$ into the heavy dark sector Higgs fields is the $T_1T_{-1}$ mode already mentioned above; even in this case, we find ourselves to be {\it extremely} close to the 
kinematic threshold resulting in a suppressed cross section. 
Of course once $\sigma B(Z_I\to l^+l^-)$ is known in the NWA, the corresponding signal rate for the $Z_I \to T_1T_{-1}$ decay process can be easily determined as the ratio of their branching 
fractions which is just given by $(T_{3I}-x_IQ_D)^2 \beta_{T_1}^3/2$, where $x_I=x_w$, $\beta_{T_1}^2=1-4m_{T_1}^2/M_{Z_I}^2$, and where $Q_D(T_1)=1, T_{3I}(T_1)=0$. Employing the 
values of our toy benchmark model parameters, this ratio is found to be $\simeq 3.84\cdot 10^{-4}$, which is indeed seen to be quite highly suppressed due to both the small coupling factor as well 
as the very highly restricted amount of available phase space. Clearly, to make use of this production mechanism for any analysis, the $Z_I$ would need to be significantly lighter than the 
expected FCC-hh search reach. A quick analysis employing Fig.~\ref{fig3} would indicate that having a sufficient number of $T_1T_{-1}$ events for 
further study would only be possible for $Z_I$ masses below roughly $\simeq 9$ TeV, corresponding to a dark vev of $v_D \lsim 24$ TeV, a region very easily accessible to the dilepton channel 
searches at FCC-hh. Thus at least we certainly would know of the existence of the $Z_I$ long before any study of the decays of the $T_1T_{-1}$ system could be performed. 

Once the conjugate pair of $T_1$ states are produced, how would they decay? No on-shell 2-body decays via the gauge bosons are seen to be kinematically allowed for $T_1$ and in the scalar 
potential, $T_1$ always appears in the combination $T_1T_{-1}$ (as it should since $Q_D$ is conserved in the limit that $v_T\to 0$) so $T_1$ cannot decay through any of the quartic terms. Thus 
we are left to consider off-shell modes such as $T_1\to h_DW_I^*$ and then one must 
address the issue of what decay modes of the (virtual) $W_I^*$ may be kinematically accessible. In previously considered scenarios\cite{Rizzo:2022qan}, $W_I$ was found to decay into 
$\bar fF$ where $f$ is a SM fermion and $F$ is a corresponding PM state with which it shares an $SU(2)_I$ doublet. $F$ may or may not be a color triplet, \ie, $\bar fF=e^+E^-,\bar dD$, in the 
obvious language employed in Ref.\cite{Rueter:2019wdf}, connecting directly to the previous discussion above. Current experimental bounds from the LHC, summarized in Ref.\cite{Rizzo:2022qan}, 
are found to be weaker for the lepton-like, color singlet $E$ PM state (just below $\sim 1$ TeV) than for the color triplet (which is roughly $\sim1.8$ TeV) so it more likely that the $W_I^*\to e^+E-$ 
path would be dominant provided that $m_E\lsim 0.15 v_D$, which is $\simeq 2$ TeV when $v_D=13.5$ TeV. Now $E^-$ itself essentially only decays to $e^-V$ so that this cascade produces 
$T_1\to h_D(\to 2V) W_I^*,W_I^*\to e^+E^-,E^-\to e^-V$, \ie, the final state $e^+e^-$ plus MET, assuming that $V$ decays to $\chi$'s. $T_1T_{-1}$ pair production thus likely yields a 4 lepton 
plus MET final state. If the $\bar d D$ were also to be accessible, the additional final states such as 4 jets plus MET or the mixed 2 jets and 2 leptons plus MET final states would also be relevant. 
Unfortunately, the relative masses of the PM fields are not fixed under our sample benchmark assumptions. 

\begin{figure}[htbp]
\centerline{\includegraphics[width=5.0in,angle=0]{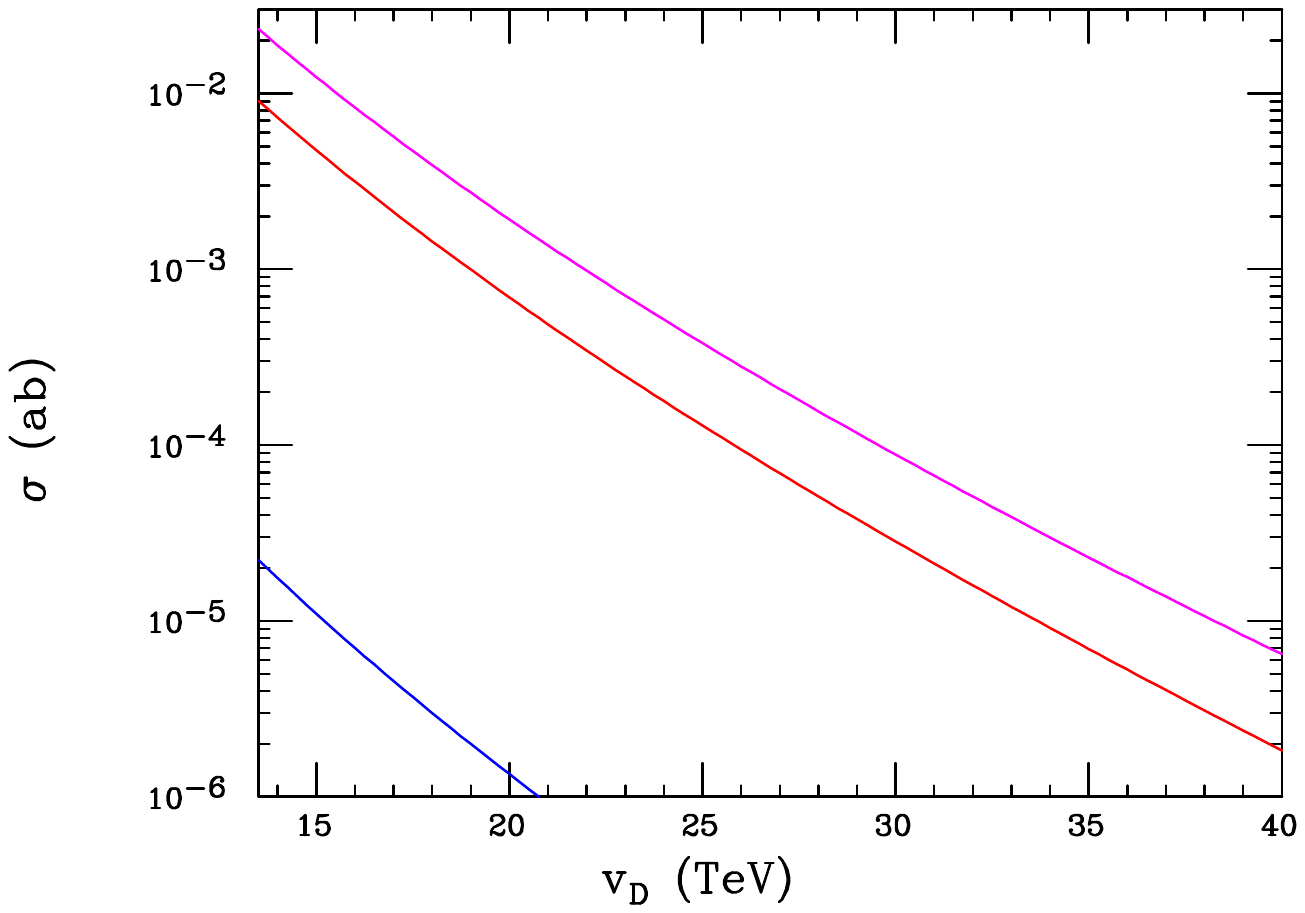}}
\vspace*{-1.3cm}
\caption{The cross sections for the $Z_Ih_1$ (blue) and $h_1I_0$ (red) channels as functions of the vev $v_D$ at the 100 TeV FCC-hh for the toy benchmark model discussed in the text.  
The corresponding result for $\bar \psi \psi$ production when we assume that $m_\psi=0.2v_D$ is shown as the magenta curve.}
\label{fig4}
\end{figure}

Now it is to be noted that the mass window for the decays of the virtual $W_I^*\to \bar \psi \chi_{1,2}$ in the $T_1$ cascade decay process may also be open depending upon the mass of the 
$\psi$ - which we know in practice due to mixing is also generated by the vev $v_D$ like the PM fields and the heavy gauge bosons. This window can occur if $m_\psi$ satisfies the same constraint 
as that for $E$ above, \ie, $m_\psi \lsim 0.15 v_D$.  If so,  the $W_I^*$ might then decay invisibly at least part of the time so that $T_1$, and hence $Z_I$, would also have additional invisible decay 
modes. 

What about the Bjorken process{\cite {bj}}, the analog of which for the present setup is the 3-body process $Z_I\to h_{1,2}Z_I^*, Z_I^*\to $dileptons? First, we observe that only the decay to the $h_1$ 
is kinematically allowed while the mass ratio $m_1/m_{Z_I}\simeq 0.45$ is completely fixed as is the relevant coupling so that the branching fraction for this process relative to the purely dilepton 
mode is as well, \ie, $\simeq 1/3000$. The above Figure then tells us that in order to have even a few dozen events of this kind to examine further would then require the mass of the $Z_I$ to lie 
not too far above the current bound from the LHC. Thus it seems likely that this interesting mode would note be very useful for our purposes.

The only other final states involving pairs of heavy dark Higgs that are in principle accessible via virtual $s$-channel  $Z_I^*$ exchange are $I_0 h_{1,2}$ with the $h_1$ being the more favored mode 
due to kinematics as well as its larger coupling arising from $h_D'-R_0$ mixing. Unfortunately, without the effect of resonant enhancement, the cross section for this final state lies significantly 
below 1 ab$^{-1}$ at the FCC-hh and so is not very useful as can be seen by the red curve in Fig.~\ref{fig4}. This same Figure show us that the associated production process at FCC-hh, 
$\bar f f\to Z_Ih_1$, displayed as the blue curve, is even further suppressed due to a smaller numerical prefactor in the cross section, a reduction in phase space, as well as a smaller mixing angle.  
From this we can conclude that if the dark scale mass spectrum were only to be slightly heavier, relative to the $Z_I$, than that considered here, all the $Z_I$ mediated processes would essentially 
become invisible; this would certainly happen if $\lambda_{1,5}>\lambda_{SM}$.

A second obvious avenue is to singly produce the $Q_D=0$ heavy dark Higgs states, $h_{1,2}$, in $gg$ annihilation via a colored fermion PM, $F$,  loop analogous to the top in the SM. Given the 
values of $v_D$ and $m_F$, the Yukawa couplings of $F$ to $h_{1,2}$ are then completely determined by the values of $s_\kappa, c_\kappa${\footnote {For concreteness, here we will assume 
the existence of only a single species of colored PM fermion. For most of the parameter space the cross section will only increase if additional states are present.}}  Since all the masses 
and mixing angles are known for our toy benchmark case, we need only fix the values of $v_D>13.5$ TeV and $m_F>1.8$ TeV\cite{Rizzo:2022qan} from present searches. The first thing to realize 
is that such searches are likely to be beyond the capabilities of the HL-LHC even with an integrated luminosity of 3 ab$^{-1}$ since $v_D> 13.5$ TeV is required from the null search $Z_I$ bound as 
noted above, implying that $m_1 \gsim 2.3$ TeV. Indeed for this mass, assuming $m_F=2.5$ TeV,  we obtain only $\sigma (gg\to h_1)=0.41$ ab (with a QCD correction K-factor included), a value 
which is far too small to be useful; the corresponding $h_2$ cross section is unsurprisingly smaller be several orders of magnitude. However, this situation changes completely at the 
100 TeV FCC-hh as can be found in Fig.~\ref{fig5}, especially for the case of the lighter state, $h_1$. Here we see that for values of $v_D\lsim 31$ TeV, in the absence of cuts or final state 
branching fractions, a sample of, say, 100 $h_1$ events would be available independently of the value of $m_F$. However, if the values of $m_F$ were relatively small, this same rate would be 
obtained even if $v_D$ approached 40 TeV. As is well known, as the ratios $m_{h_{1,2}}^2/4m_F^2 \to 0$, the lowest order loop function responsible for this process goes to a constant and we 
see that predictions for the 
values of $m_F=10,30$ TeV are always rather similar and that for $m_F=2.5$ TeV joins up with these when $v_D$ get close to its lower bound. In the case of $h_2$, the larger mass for the same 
value of $v_D$ significantly reduces the production cross section relative to that of $h_1$ and we see that to get the same 100 event sample we would need to restrict $v_D$ to values $\lsim 21$ TeV.  
Still, this process seems far more advantageous than does $Z_I$ resonant production especially if $\lambda_{1,5}>\lambda_{SM}$.

\begin{figure}[htbp]
\centerline{\includegraphics[width=5.0in,angle=0]{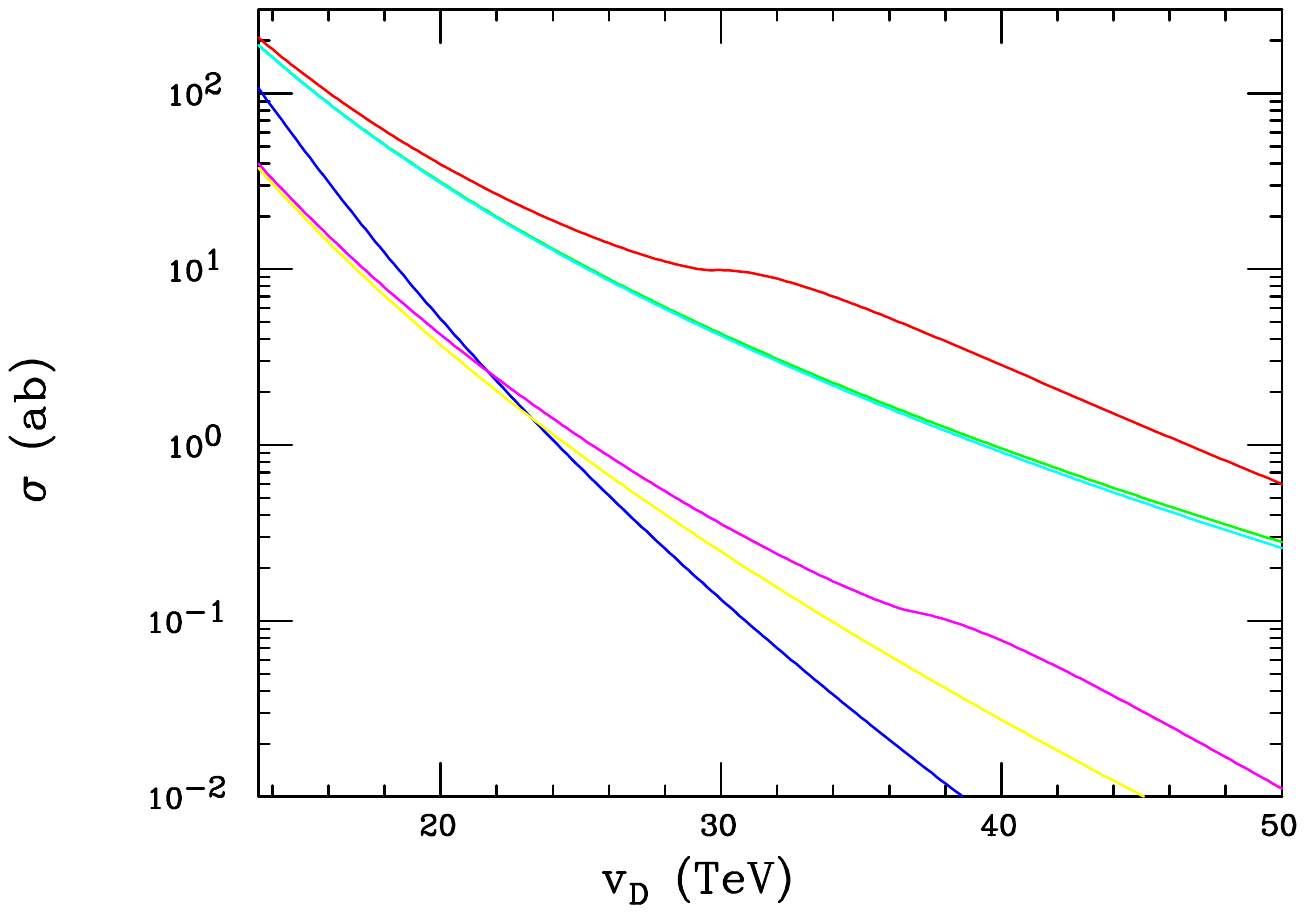}}
\vspace*{-1.3cm}
\caption{The $gg\to h_{1,2}$ production cross sections at the 100 TeV FCC-hh as a function of the vacuum expectation value, $v_D$, in our toy benchmark model assuming different values of 
the color PM fermion mass, $m_F$, running in the loop and a K-factor of 2.7. The red (green, cyan) curves are for $h_1$ production with $m_F=2.5(10,30)$ TeV, respectively. The blue (cyan yellow) 
curves are the analogous results for $h_2$. }
\label{fig5}
\end{figure}

$h_2$, being relatively massive, has many kinematically allowed 2-body decay modes via the scalar potential, \eg, $T_1T_{-1}, 2I_0, 2h_1,2h_D$, but the decays to the gauge boson pairs can only 
occur if one of then is off-shell,\ie, $W_IW_I^{\dagger *}, Z_IZ_I^*$, similar to the SM Higgs decays into $WW^*$ and $ZZ^*$. $h_1$, on the other hand, being the lightest of these heavy dark Higgs 
states, has very few 2-body decay avenues which are open to it if the final state particles are all required to be on-shell. One obviously allowed mode is $h_1\to 2h_D$ whose coupling 
$\sim \lambda_4v_D s_\kappa$ arises from $V_{Dark}$;  since the two $h_D$ decays invisibly, an additional ISR jet would be needed to produce a visible, monojet-like final state. 
An analogous decay $h_1\to 2h_{SM}\sim 2H(125)$ can also occur via the coupling $\sim \lambda' v_D s_\kappa$ arising from $V_{SM}$. However, we expect that $\lambda'$ is 
somewhat suppressed due to the mixing constraint requirements discussed previously above so that this decay would likely not have a very large branching fraction. However, for certain 
parameter space regions, this mode may provide the best $h_1$ production signature. 

For completeness we note that the FCC-hh direct search reach for the pair production of the new heavy color triplet PM fermions themselves which run in the $ggh_{1,2}$ loop will likely be 
somewhat less than roughly $\sim 10$ TeV, assuming that they dominantly decay into jets plus MET, \ie, $F\to fV$, as expected, even though they are produced via the strong interactions.
 
Other analogs of the production of Higgs bosons in the SM or the THDM are also found to lead to highly suppressed rates due to the large mass scales involved. An example of this is heavy 
gauge boson fusion\cite{heavy}, \eg, $\bar f f \to \bar FF+W_I^*(W_I^(\dagger *)\to \bar FF+h_{1,2}$, through the vev $v_D$, which requires the production of a pair of heavy PM fields in the final 
state in addition to the dark Higgs and proceeds through double $t$-channel exchanges of the $W_I$'s. The analogous double $Z_I$ exchange process is somewhat better off, even though the 
fermionic couplings are somewhat smaller, as the PM pair in the final state is now replaced with a (essentially massless) pair of SM fermions but this, unfortunately, will suffer from the double 
$Z_I$ propagator suppression. Obviously, the mixed $Z_IW_I$ fusion process is also suppressed for these same reasons.

Finally, we need to consider the production of the new $Q_D=0$, SM singlet vector-like fermion, $\psi$, that inhabits the $SU(2)_I$ doublet with the DM $\chi$. Needless to say, this state is clearly 
very difficult to produce as its mixing with, \eg, the SM Dirac neutrino is quite infinitesimal. From the discussion above, it is also clear that the most optimistic possibility is resonant $Z_I$ production 
followed by the on-shell decay to $\bar \psi \psi$, this then requiring that the $\psi$ mass satisfy the constraint $m_\psi \lsim 0.185v_D$. In such a case, we find that  
$\Gamma(\bar \psi \psi)/\Gamma(e^+e^-)=\beta_\psi (3-\beta^2_\psi)$ with $\beta^2_\psi=1-4m_\psi^2/m_{Z_I}^2)$, which is relatively close to unity except near the kinematic boundary, and so 
the production rate can be read off directly from Fig.~\ref{fig3}. If $m_\psi$ exceeds this bound, then the $Z_I$ will again be off-shell and the corresponding production cross section becomes highly 
suppressed and invisibly small as was the case for the production of heavy dark Higgs pairs above. For example, if $m_\psi=0.2v_D$, then the cross section would only be a few times larger than 
that obtained for $h_1I_0$ as seen in Fig.~\ref{fig3} and which is still too small to be useful.  If $\psi$ does satisfy this mass bound, once it is produced, $\psi$ will decay via a virtual 
$W_I$, \ie, $\psi \to \chi W_I^*$, and then one is returned to the problem already encountered above of ascertaining how this off-shell $W_I$ would subsequently decay. 

Since the SM quartic coupling, $\lambda_{SM}$, is relatively small, it is clear that if the actual $\lambda_i$ in this framework are, as a whole, somewhat larger then the entire dark scalar spectrum 
could be pushed into the region of inaccessibility at colliders due to their weak couplings to the visible sector other than possibly by single production via the $gg$-fusion mechanism.

\section{Discussion and Conclusion}

Portal Matter induced kinetic mixing between the $U(1)_D$ dark photon and the SM gauge fields at the 1-loop level offers an attractive picture for light thermal dark matter in the mass range below 
$\sim 1$ GeV, allowing the observed relic density to be achieved in a manner consistent with all other experimental constraints. In models of this type, the SM singlet DM carries a dark charge, 
$Q_D=1$, under this $U(1)_D$ while all the SM fields have $Q_D=0$ and thus an interaction of SM particles with the dark sector is only generated by KM. In such a setup, data from the CMB 
and from other astrophysical/cosmological observations tells us that the DM in this mass range must have a suppressed annihilation rate into SM fields at later times. This implies that, if it is 
fermionic, the DM must be either a Majorana or pseudo-Dirac state so that $p$-wave and/or Boltzmann suppression of the annihilation rate(s) will occur at these later times. Simultaneously, 
it has been shown that the required low energy content of such a setup may force the $U(1)_D$ gauge coupling, over much of its phenomenologically interesting range, to run by RGE evolution 
into the non-perturbative regime at or before the scale of $\sim10$'s of TeV unless a new, asymptotically-free, non-Abelian UV-completion, $G_D$, occurs. The breaking of $G_D\to U(1)_D$ then 
gives masses to both the set of additional gauge bosons beyond the DP as well as to the PM fields. Within such a setup, which has been the subject of much of our previous work, the simplest 
possibility is that $G_D=SU(2)_I\times U(1)_{Y_I}$, in analogy with the SM, while also taking the DM to be a complex, vector-like singlet under this group having a dark charge. Then, once the light 
dark Higgs coupling to this state generates the required Majorana mass terms, the original Dirac complex singlet DM state splits into two distinct eigenstates whose couplings to the DP are 
controlled by the mixing angle required to diagonalize the corresponding $2\times 2$ mass matrix; this has a direct impact on the calculation of the relic density and also the cross section relevant 
for DM direct detection searches.  But one might ask how this somewhat simple picture is altered if we no longer make the simplifying assumption that the DM is a $G_D$ singlet as this will certainly 
not be the case in a more general model.

In this paper, we addressed this question by considering the next simplest possibility wherein the DM, $\chi$, instead lies in an $SU(2)_I$ doublet together with another SM singlet, vector-like fermion, 
$\psi$,  which has $Q_D=0$. In addition to this new fermionic state, the $Q_D=2$ dark Higgs field whose vev is necessary to generate the required Majorana mass terms for the DM (as well as the 
mass for the DP) must now reside as part of an $SU(2)_I$ triplet implying the existence of several additional dark Higgs fields not previously encountered. We then ask whether this augmentation 
of the fermion and scalar sectors leads to some new physics.  In performing this study, we recall that there are three widely separated physics scales in these types of setups arising from the three 
hierarchal vevs which are responsible for the breaking of the various gauge symmetries and the generation of particle masses: ($i$) the very large scale, $v_D\sim 10$'s of TeV, at which $G_D$ 
breaks down to $U(1)_D$ and which generates the PM masses as well as those for most of the dark Higgs field content of the model. Here, specifically, $v_D$ is the vev of an $SU(2)_I$ doublet. 
($ii$) Of course the usual SM doublet vev, $v_{SM} \simeq 246$ GeV needs to be present; and finally ($iii$) the triplet vev corresponding to the scale at which $U(1)_D$ breaks and the DP and 
DM obtain their masses, $v_T\lsim 1$ GeV.  In such a case it is quite useful to examine the limit wherein the ratios of (the squares of ) these vevs are treated as very small parameters. We further 
recall that in the $v_T\to 0$ limit, $U(1)_D$ remains unbroken and so $Q_D$ will remain a conserved quantity. Since these new states have masses set by $v_D$ and are rather weakly coupled, 
many, is not all, of them are forced lie beyond the reach of the LHC and so it will require the 100 TeV FCC-hh to provide any possible direct access for their production and examination.

These extensions to the dark sector field content lead to some interesting new phenomenological implication beyond those already encountered in the complex singlet DM version of this 
setup. The details are, however, found to be a bit sensitive to the mass spectrum of these new states. 
Note that although the light $|Q_D|=2$ dark Higgs generating the Majorana mass terms for the DM now lies in a triplet representation (instead of another singlet) of $SU(2)_I$, the DM 
phenomenology at low energies, \ie, its mix of couplings to the DP, \etc, goes through essentially as in the case of singlet DM. However, ($i$) the existence of the $Q_D=0$, SM singlet Dirac fermion, 
$\psi$, as part of the $SU(2)_I$ doublet where the DM resides and nominally with the same Dirac mass as the DM leads to a significant suppression of the `expected' value of the (Dirac) neutrino 
mass in the SM, by a factor of order $\sim 10^5$. This occurs due to a coupling between $\psi_L$ and the (also SM singlet) right-handed neutrino, $\nu_R$, that is induced by the $\sim 10$'s of 
TeV scale, $SU(2)_I$-breaking doublet vev, $v_D$, simultaneously generating a much larger mass for $\psi$. Having a mass set by $v_D$, the new mass eigenstate (almost purely $\psi$) can 
make itself directly felt only in collider searches. 

In the $v_T/v_D\to 0$ limit, as noted, $Q_D$ remains conserved, making it difficult to produce and study the new dark states at colliders, especially due to the plethora of model parameters. To this 
end we chose to study a particular benchmark point wherein these new parameters took on their corresponding SM values leaving only the scale set by $v_D$ as free; this is a relatively optimistic 
choice. In such a case, the entire spectrum of new dark states will have fixed {\it relative} masses and existing LHC searches at 13 TeV for the heavy hermitian gauge boson, $Z_I$, in the dilepton 
mode at the LHC tell us that $v_D\gsim 13.5$ TeV, placing a lower bound on this spectrum. This lower bound is sufficient to inform us that the FCC-hh is required to pair produce these dark states 
via on- or off-shell $Z_I$ exchange or to make either of the two, $Q_D=0$,  scalar states singly in $gg$-fusion via a colored PM loop. The rate for this later process is found to be rather respectable 
for a wide range of both $v_D$ and PM masses as the single production of a heavy scalar saves us from paying a potentially very large phase suppression as is seen in Fig.~\ref{fig5}.  At least for 
the lighter state, $h_1$, since it would most dominantly decay invisibly, an additional ISR jet would be required to tag these types of production events. In the case of scalar pairs, only the 
$Q_D=1$ state, $T_1$, can be pair produced by resonant, on-shell $Z_I$ decay due to the nature of the scalar spectrum, but it still suffers from having both a small couplings to $Z_I$ and the fact 
that it is a $p-$wave process with $2m_{T_1}\simeq m_{Z_I}$ so that it is highly phase space suppressed. In comparison to dilepton production, this on-shell mode is found to be suppressed by a 
relative factor of $\sim 2500$ so, 
as can be seen from Fig.~\ref{fig3}, it can only be studied for values of $v_D$ not very much larger than the current LHC bounds. Unfortunately, the simplifying assumptions made above do not 
impact the PM spectrum and their masses are in partial control of the possible final states that will occur in $T_1$ decays, as well as those of the other new scalars. With the remaining 
uncertainties in the PM spectrum, the most likely final state for $T_1T_{-1}$ production would be 4 charged lepton plus MET which takes place through intermediary $W_I$ and the lepton-like PM 
fields. The mass spectrum of the remaining dark scalars is such that other 
pair production modes via the $Z_I$ necessarily takes place off-shell and so are quite significantly suppressed with rates too small to be useful as can be seen in Fig.~\ref{fig4}.

The $Q_D=0$ heavy fermion $\psi$ {\it might} be accessible on-shell in $Z_I$ decay provided that it is sufficiently light in which case it would have a significant pair production rate comparable to 
that of the dilepton mode. If, on the other hand, $2m_\psi > m_{Z_I}$, then the rate is again found to be too small to be observable. Once produced, $\psi$ will then decay to $\chi W_I^*$, with the 
off-shell $W_I$ itself most likely decaying into dileptons plus MET but, as before, this will depend upon the unconstrained details of the PM mass spectrum.

The KM picture of light thermal dark matter is very appealing but will require some variety of UV-completion. Hopefully signals of both dark matter and dark photons will soon be experimentally 
observed.

\section*{Acknowledgements}
The author would like to particularly thank J.L. Hewett for valuable many discussions and the Brookhaven National Laboratory Particle Theory Group for its hospitality.  This work was supported 
by the Department of Energy, Contract DE-AC02-76SF00515.



\end{document}